\documentstyle[12pt,epsf]{article}

\begin{document}
\vspace*{0.5cm}
\centerline{PACS 03.65.-w; 21.10.Hw; 21.10.Ky}
\centerline{Preprint IBR-TH-=97-039, October 15, 1996}

\bibliographystyle{unsrt}
\newcommand{\Po}{Poincar\'e}

\def\spin{\mbox{\scriptsize\rm spin}}
\def\orb{\mbox{\scriptsize\rm orb}}
\def\tot{\mbox{\scriptsize\rm tot}}
\def\diag{\mbox{\rm diag}}
\def\rmexp{\mbox{\scriptsize\rm exp}}
\def\theor{\mbox{\scriptsize\rm theor}}
\def\res{\mbox{\scriptsize\rm res}}
\def\stim{\mbox{\scriptsize\rm stim}}
\def\spont{\mbox{\scriptsize\rm spont}}
\def\D{\mbox{\scriptsize\rm D}}
\def\QM{\mbox{\scriptsize\rm QM}}
\def\c{^{\circ}}
\def\min{\mbox{\scriptsize\rm min}}
\def\max{\mbox{\scriptsize\rm max}}
\newcommand{\tim}{\times}
\def\mdexp{\mu_{\mbox{\scriptsize\rm D}}^{\mbox{\scriptsize\rm exp}}}
\newcommand{\no}{\noindent}
\def\h{\hat}
\def\arr{\rightarrow}
\def\dag{\dagger}
\def\l{\langle}
\def\r{\rangle}
\def\cc{{\cal C}}

\title{\large\bf Use of relativistic hadronic mechanics for the exact
representation of nuclear magnetic moments and the prediction of new
recycling of nuclear waste}
\author{Ruggero Maria Santilli\\*[1ex]
{\normalsize\it Institute per la Ricerca di Base}\\
{\normalsize\it Castello Principe Pignatelli, Molise, Italy}}
\date{}
\maketitle

{\bf Summary.} We present a new realization of relativistic hadronic
mechanics and its underlying iso-{\Po} symmetry specifically constructed
for nuclear physics which: 1) permits the representation of nucleons as
extended, nonspherical and deformable charge distributions with alterable
magnetic moments yet conventional angular momentum and spin; 2) results to
be a nonunitary ``completion'' of relativistic quantum mechanics much along
the EPR argument; yet 3) is axiom-preserving, thus preserves conventional
quantum laws and the axioms of the special relativity. We show that the
proposed new formalism permits the apparently first exact representation of
the total magnetic moments of new-body nuclei under conventional physical
laws. We then point out that, if experimentally confirmed the alterability
of the intrinsic characteristics of nucleons would imply new forms of
recycling nuclear waste by the nuclear power plants in their own site,
thus avoiding its transportation and storage in a (yet unidentified)
dumping area. A number of possible, additional basic advances are also
indicated, such as: new understanding of nuclear forces with nowel
nonlinear, nonlocal and nonunitary terms due to mutual penetrations of the
hyperdense nucleons; consequential new models of nuclear structures;
new magnetic confinement of the controlled fusion  taking into account
the possible alterability of the intrinsic magnetic moments of nucleons at
the initiation of the fusion  process; new sources of energy based on
subnuclear processes; and other possible advances. The paper ends with
the proposal of three experiments, all essential for the continuation of
scientific studies and all of basic character, relatively moderate cost and
full feasibility in any nuclear physical laboratory.


\section{Open character of total nuclear\hfil\break magnetic moments.}

One of the fundamental, still unsolved problems of contemporary nuclear
physics is the exact representation of the total magnetic moments of
nuclei, particularly those of few-body nuclei such as the deuteron, tritium
and helium in view of their known limited number of free parameters.

As an example, the experimental value of the deuteron magnetic moment is
given by

$$
\mdexp =0.857406\:,\eqno{(1.1)}
$$
\def\no{\noindent}

\no while its representation via nonrelativistic quantum mechanics (QM)
on D-states yields the familiar epression (see, e.g., [1]).

$$
{\mu_{\QM}}^{\theor}=g_N+g_p=0.880\:,\eqno{(1.2)}
$$

\no (where $g_p$ and $g_n$ are the gyromagnetic factors of the protons and
neutron, respectively) which is about 26 \% off {\sl in excess} of the
experimental value.

It is known that the use of all possible correction via relativistic
quantum mechanics (RQM) on a mixture of $S$-, $D$- and $P$-states can
reduce the above deviation down to about 1 \%, but RQM {\sl cannot}
produce an exact representation of $\mdexp$, as confirmed
by the recent studies [2].

It is equally known that the remaining 1 \% deviation cannot be eliminated
via quark theories because, unlike the corresponding case in the atomic
structure, the quark orbits are very small, and their polarization yields
corrections to the total magnetic moment of nucleons of the order of
$10^{-3}$ \%.

A similar situation exists for representation of the total magnetic moment
of the tritium helium and other few-body nuclei.

The most plausible explanation of the above occurrence was formulated by
the Founding Father of nuclear physics in the late 1940's immediately after
the identification of the numerical value  (1.1). For instance, in p. 31 of
[1] one can read: {\sl ``It is possible that the intrinsic magnetism of
nucleon is different when it is in close proximity to another nucleon.''}

Recall that nucleons are not point like, but have an extended charge
distribution with the radius of about 1 fm ($10^{-13}$ cm). Since perfectly
rigid bodies do not exist in the universe, it is plausible to expect that
such distribution can be deformed under sufficient external forces. But the
deformation of a charged and spinning sphere implies a necessary alteration
of its intrinsic magnetic moment, as established by classical electrodynamics.

The above ``historical hypothesis'' (as referred to hereon) therefore
assumes that, when a proton and a neutron are coupled into the deuteron
or other nuclear structure, their charge distributions are altered by the
nuclear force, resulting in an alteration of their conventional
{\it intrinsic} magnetic moments as measured in vacuum. In turn the
assumption of a departure from standard quantum values of the magnetic
moments of nucleons readily permits an exact representation of the total
magnetic moment of few-body nuclei, as we shall see in Sect. 4.

Since no exact representation of $\mdexp$ has been achieved
via {\sl conventional} intrinsic magnetic moments of nucleons following
about three-quarter of a century of attempts, $\mdexp$ should
be assumed to constitute {\sl direct experimental evidence on the alterability
of the intrinsic magnetic moments of nucleons in the transition from motion
in vacuum to motion within nuclear structure.} Note that the representation
of $\mdexp$ requires a {\sl decrease} of the intrinsic magnetic
moments of nucleons, and that such a decrease can only occur for a
{\it prolate} deformation of nucleons referred to their spin axis.

\vspace{1cm}

FIGURE 1. {\bf The historical hypothesis on the alterability of the intrinsic
magnetic moments of nucleons.}
A schematic view of the historical hypothesis on the deformability
of the intrinsic magnetic moments of nucleons which constitutes the main topic
of study of this paper. As we hope to illustrate, the above hypthesis may
stimulate a new scientific renaissance in nuclear physics because it is
beyound realistic means of quantitative representation via the Poincar\'e
symmetry and RQM, thus requiring their structural generalizations with far
reaching implications of conceptual, theoretical, experimental and
applicative character.

\vspace{1cm}

Note finally that the above historical hypothesis is {\sl model independent},
i.e., it applies independently from any assumed structure of the nucleons,
and consists of the geometric deformation of their charge distributions
whatever the constituents are.

Additional preliminary experimental evidence on the alterability of the
intrinsic magnetic moments of nucleons were conducted from 1975 to 1979
by H. Rauch and his associates [3] via interferometric measures of the
$4\pi$ {\it spinorial symmetry} of the neutron. The measures are performed
via a familiar perfect crystal which splits a thermal neutron beam into two
branches which are then coherently recombined. In one (or both) branches
experiments [3] placed an electromagnet calibrated at 7,496 G which, for the
{\sl conventional} value of the intrinsic magnetic moment of the neutron,
would yield an exact multiple of two complete ($4\pi =720^{\circ}$) spin
flips, as requested by the Fermi-Dirac character of the neutron and as
necessary for a coherent recombination.

In order to improve accuracy, the experimenters filled up the electromagnet
gap with Mu-metal sheets which reduce stray fields [3]. While crossing the
electromagnet gap, the neutron beam is therefore exposed to the field of
7,496 G as well as to the intense electric and magnetic fields in the
vicinity of Mu-metal nuclei. The best interferometric measures date back
to 1979 with re-elaboration done in 1981 [3e], and are given by

\def\c{^{\circ}}

$$
\theta=715.37^{\circ}\pm 3.8^{\circ}\:,\quad\theta_{\min}=
712.07\c\:,\quad\theta_{\max}=719.67\c\:.\eqno{(1.3)}
$$

\no Such, they {\sl do not } contain $720\c$ in the minimal and maximal
values. However, the deviation is {\sl smaller} than the error and,
therefore, the above measures are inconclusive.

Similar measures were conducted in 1975 by S.A. Werner and his associates
[3f] although also with unsettled results. To our best knowledge, no
additional interferometric measures have been done for the $4\pi$ spinorial
symmetry of the neutron since 1979, thus indicating the need for final
tests which are now permitted in view of the technological advances and
improved accuracy occurred since the late 1970's.

\def\tim{\times}

Despite the above unsettled character, measures (1.3) are significant, as
shown in theoretical studies [4]. In this respect let us recall the
$(p,q)$-{\it deformations} of Lie algebras first introduced by Santilli [4a]
back in 1967 as part of his Ph.D. studies with product
$(A,B)=p\tim A\tim B-q\tim B\tim A$ of Albert's [4a] {\it Lie-admissible}
type, where $p$, $q$, and $p\pm q$ are nonzero parameters and $A\tim B$ is
the usual associative product.

By using the preceding deformations Eder [4c,4d] has shown that the
alteration of the charge distribution of the neutron caused by the intense
{\it electric} and {\it magnetic} fields in the vicinity of Mu-metal nuclei
could indeed yield ``spin fluctuations'' with about 1 \% deviation of the
intrinsic magnetic moment which is precisely in order of magnitude needed
for the resolution of the historical problem of total nuclear magnetic
moments. Note that the {\it strong interactions} of Mu-metal nuclei have
an irrelevant conribution here because their sectional area along the
thermal neutron beam is very small.

Also, all {\it median} angles measured in tests [3] (with the electromagnet
gap filled up with Mu-metal sheets) are {\sl smaller} than the expected
$720\c$. This occurrence was studied by Santilli [4e] via the first
$(p,q)$-deformations of the SU(2) spin algebra and called
{\it angle slow-down effect}. This apparent effect is significant inasmuch
as it requires a {\sl decrease} of the standard magnetic moment of the
neutron for the (polarized) conditions of the experimental set up which is
precisely in line with the {\sl decrease} of the same magnetic moments needed
for the interpretation of $\mdexp$, as recalled earlier.

The electric and magnetic fields in the vicinity of Mu-metal nuclei are
known and result to be of order (in average) of 20,000 G. The biggest
unknown is the {\sl deformability of the charge distribution of neutrons}
under known external fields, which can only be established from
$\mdexp$ for the deuteron conditions are done in Sect.4,
or via interferometric measures for more general conditions.

As noted earlier, $4\pi$-interferometric tests can only measure the
deformability of neutrons under the intense {\it electric} and {\it magnetic}
fields of the Mu-metal (or other heavy) nuclei, but not under the {\it strong
nuclear forces} as occuring in the structure of the deuteron.

However, it is known from classical electrodynamics that a small deformation
of a spinning and charged sphere can yield a relatively large change of its
magnetic moment. Also, the deformability of the charge distributions
of nucleons in the deuteron structure may eventually be due to the electric
and magnetic fields of the nucleons themselves. Intense electric and
magnetic fields of {\sl large, many-body nuclei} could therefore approximate
sufficiently well the  electric and magnetic fields of the {\sl two-body}
deuteron.

The above aspects, combined with the resolution of the still open historical
hypothesis as well as with its implications pointed out in Sect. 5, are
sufficient to warrant the study of novel methods for the {\sl exact}
representation of $\mdexp$ , as well as the finalization of interferometric
measures on the $4\pi$-spinorial symmetry of the neutron.

\section{Expected lack of exact character of quantum mechanics for the
nuclear structure}

QM is fully established as being {\sl exactly} valid for the so-called
{\it exterior problems}, here referred to as particles moving in vacuum
under action-at-a-distance/potential interactions at sufficiently large
mutual distances to allow an effective point-like approximation of their
wavepackets and/or charge distributions, as occurring in the atomic
structure and electroweak interactions at large. In fact, QM provided an
{\sl exact} representation of {\sl all} experimental data available for the
systems considered.

\vspace{1cm}


FIGURE 2. {\bf Experimental insufficiencies of quantum mechanics in nuclear
physics}.
QM is {\sl exactly} valid in the atomic structure because it
provided an exact representation and understanding of {\sl all} its
experimental data. On the same grounds, QM cannot be {\sl exactly} valid for
the nuclear structure, because it has been unable to provide an {\sl exact}
representation of various experimental data. For instance, total nuclear
magnetic moments do not follow QM predictions, but are within minimal and
maximal values reproduced in this figure from [1] which motivated the
historical hypothesis of Fig. 1. Additional insufficiencies exist for:
nuclear forces; nuclear structures; total angular momenta; and other aspects.
Needless to say, QM provides an excellent approximation of nuclear data.
We are therefore referring to deviations which are generally small, yet they
have rather important implications, as indicated in Sects 4 and 5.

\vspace{1cm}

By comparison, QM {\sl is not} expected to be exactly valid for the so-called
{\it interior problems}, here referred to particles whose wavepackets or
charge distributions cannot be effectively approximated as being point-like
because moving at small mutual distances (of the order of 1 fm), as occurring
in the structure of nuclei (as well as of hadrons and stars not considered
in this paper).

The understanding is that the {\sl approximate} validity of QM in nuclear
physics is out of scientific debate. We are therefore referring to expected
{\sl small deviations} from QM treatments.

The reasons for the above occurrence are numerous. First, unlike the
corresponding atomic case, QM has been {\sl unable} to provide an
{\sl exact} representation of {\sl all} nuclear experimental data. The lack
of exact representation of total nuclear magnetic moments considered
in Sect. 1. is only {\sl one} of several isufficiencies. As an example, Ref.
[1] indicates the existence of additional lack of final understanding of:
nuclear structures, total nuclear angular momenta, and other aspects.

The above experimental insufficiencies can be established in a rigorous
theoretical way via primitive symmetry principles. Computer visualizations of
the fundamental summetry of QM, the Galilean symmetry G(3.1) or the
{\Po} symmetry P(3.1), establish their {\sl exact} validity for
{\it Keplerian systems}, that is, for systems of particles without collisions
admitting their heaviest element in the center (Keplerian center). This
confirms the exact character of QM for the atomic structure.

By comparison, {\sl nuclei do not possess nuclei} and, consequently, the
Galilei and {\Po} symmetries {\sl cannot} be exact for the nuclear
structure. In fact, the lack of Keplerian center
requires a necessary breaking of the above symmetries. In turn, any
expectation of achieving via QM an {\sl exact} representation of {\sl all}
nuclear data under these conditions has no theoretical ground.

Not surprisingly, the latter aspects are deeply linked to the preceding ones.
In fact, according to the Galilean and {\Po} symmetries in their conventional
realization (see Sect. 3 for a more general realization) {\sl the intrinsic
magnetic moment of nucleons is perennial and immutable.} Any quantitative
representation of the historical hypothesis of their deformability therefore
requires a {\sl necessary} deviation from the above symmetries, thus
confirming the mutual compatibility of the two aspects.

Both preceding aspects can be rigorously established on dynamical grounds.
QM was established for the characterization of action-at-a-distance
interactions solely derivable from a potential and this confirms again its
exact validity for the atomic structure, this time on dynamical grounds.

By comparison, nucleons in a nuclear structure are in an average state of
mutual penetration of about $10^{-3}$ parts of their charge distribution [4f].
But hadrons are some of the densest objects measured in a laboratory until
now. This indicates the presence in the nuclear force  of interactions which
are: 1) of {\it contact}, i.e., of zero-range type; 2) {\it nonlinear} in the
wavefunctions and possibly their derivatives; 3) {\it nonlocal} of a type
requring an integral over the volume of overlapping; 4) {\it nonpotential} in
the sense of violating the conditions to be derivable
from a potential or a Hamiltonian; and 5) of cosequential {\it nonunitary}
type. By recalling the strictly action-at-a-distance, linear,
local-differential, Hamiltonian and unitary character of QM, the preceding
characteristics of the nuclear force due to mutual penetration of the
hyperdense charge distribution of nucleons are dramatically beyond
any hope of quantitative QM treatment.

It should be stressed again that the above isufficiencies {\sl cannot}
be resolved via the transition to quark theories on numerous, independent,
experimental and theoretical grounds. Besides their inability to achieve
the needed exact representation of {\sl all} nuclear data, the current
theories on the hadronic structure are also of action-at-a-distance, linear,
local-differential, Hamiltonian  and unitary character, thus being unable
to represent the above expected characteristics of the nuclear force.

\vspace{1cm}


FIGURE 3. {\bf Theoretical insufficiences of quantum mechanics in nuclear
phusics.} An illustration of the theoretical impossibility for QM to be
{\sl exactly} valid for the nuclear structure due to its lack of Keplerian
center which requires a necessary breaking of the Galilean and {\Po}
symmetries. In turn, the above occurrence is only a consequence of the
fundamental theoretical insufficiencies of QM to represent nucleons as
{\sl extended, nonspherical and deformable} charge distribution, as well as
the inability to represent the component in the nuclear force expected from
their mutual penetration which is of {\it contact, nonlinear, nonlocal,
nonhamiltonian and nonunitary} type. An axiom-preserving broadening of  QM
and its underlying symmetries capable of providing a quantitative
representation of the above characteristics is outlined in Sect. 3 and
applied in Sect. 4.

\vspace{1cm}

Also, quark theories in their conventional formulation are affected by still
unresolved basic problems, such as: the lack of a rigorous confinement of the
unobservable quarks as prohibited by  Heisenberg's uncertainty principle;
the inability of quarks to be a representation of the {\Po} symmetry, thus
preventing their mathematical parameters called ``masses'' from being
rigorously defined in our space-time (as the eigenvalues) of the second-order
Casimir invariant of P(3.1)); the complete lack of gravity for any nucleus
assumed to be made up of quarks because of the impossibility of defining
gravity in current quark theories (gravity is solely defined in our
spce-time while quarks are solely defined in a mathematical unitary space
without interconnections due to the  O'Raifeartaigh theorem or known
resolution via supersymmetric models).

Thus, any attempts at shifting open problems in our current description of
nucleons in our space-time to other, considerably more serious, open
problems in their quark constituents, is a {\sl de facto} abandonment of
the search for a deeper understanding of the nuclear force and structure.

This leaves no other choice than the search, conducted in Sect. 3, of
a broadening-covering of QM capable of providing a quantitative representation
of the nuclear aspects under consideration.

The above aspects can be best illustrated via open problem of the total
magnetic moments of few-body nuclei. In fact, any quantitative study of
the historical hypothesis herein considered requires the introduction of the
following {\sl new notions:}

1) The {\sl extended, nonspherical and deformable shape} of the charge
distribution of nucleons, expectedly of spheroidal ellipsoidic character,
hereon represented with the quantities ${n_1}^2,{n_2}^2,{n_3}^2,n_k\neq 0,\:
k=1,2,3\,,$ which are functions of intensity of external fields and any other
needed local2 characteristic. For particles with spin along the third axis,
the above quantities represent a spheroidal ellipsoids which are {\it oblate}
for ${n_1}^2={n_2}^2>{n_3}^2$ and {\it prolate} for ${n_1}^2={n_2}^2<{n_3}^2$.
The evident condition of preserving the original volume of nucleons then
yields the normalization hereon assumed

$$
{n_1}^2\tim {n_2}^2\tim {n_3}^2=1\:,\quad{n_1}^2={n_2}^2>\mbox{or}<{n_3}^2\:.
\eqno{(2.1)}
$$

\no [It should be noted that in other cases the  normalization
${n_1}^2+{n_2}^2+{n_3}^2=3$ may be preferable].

2) The {\it density} of the medium in which motion occurs hereon represented
with the functions ${n_4}^2$ which, for the vacuum, is assumed to have the
normalized value ${n_4}^2=1$, and  we shall write

$$
{n_4}^2=1\:,\quad < 1\:,\:\:\mbox{or} > 1\:.\eqno{(2.2)}
$$

\no As we shall see in Sect. 3, $n_4$ is in reality the local {\it index
of refraction of light}, thus characterizing the local causal speed.

3) The alteration called {\it mutation} [6b] of the intrinsic magnetic
moment $\mu_N$ of nucleons $N=n,p$, hereon expressed with the symbol
$\hat{\mu}_N=\hat{\mu}_N(\mu_N,\\{n_{\mu}}^2,\ldots)$, $\mu=1,2,3,4$, with
$\hat{\mu}_N> \mu_N$ for {\it oblate} deformations and $\hat{\mu}_N< \mu_N$
for {\it prolate} ones, where the term ``mutation'' is preferred over
``deformation'' to indicate the fact that the underlying methods [6]
(see the next sections) are structurally different than the known
``quantum deformations'' [4] of the current literature.

It is evident that the nonspherical and deformable characteristics
(2.1) are beyond any representational capability of QM because the latter
can only represent {\sl perfectly spherical and perfectly rigid particles},
as necessary in order not to violate the fundamental rotational symmetry
O(3). It should be stressed that the same occurrence persists in
second-quantization and related form-factors which cannot represent the main
characteristics of the historical hypothesis under study here. By comparison,
any real treatment of the historical hypothesis requires {\it ab initio}
the representation of {\sl nonspherical and deformable particles}.

The above limitations of QM are well known  to be inherent in the very
structure of its fundamental carrier spaces, the Euclidean space
$E=E(r,\delta ,{\cal R})$ with coordinates $r=\{r^k\},\:k=1,2,3$, and metric
$\delta =\diag\, (1,1,1)$ over the field of real numbers
${\cal R}={\cal R}(n,+,\tim)$ and the Minkowski space $M=M(x,\eta ,{\cal R})$
with coordinates
$x=\{x^{\mu}\}\,,\:\mu =1,2,3,4$, and metric
$\eta =\diag\, (+1,+1,+1,-1)$
over ${\cal R}\:.$ In fact, the basic unit of $E$, $I=\diag\,\{1,1,1\}$
(which is the space component of the unit of $M$) represents a
{\sl perfect and rigid sphere}. Moreover, the theory of deformations is
well known to be incompatible with the above spaces, their symmetries
and, consequently, QM.

Alternatively, it is easy to see that deviations from the exact $720\c$
in the $4\pi$ interferometric measures (1.3) imply a deviation from the
familiar {\it spinorial} component of  Dirac's wavefunction,

$$
\psi '=R(\theta_3)\tim\psi =e^{i\gamma_1\gamma_2\theta_3/2}\tim\psi\:,
$$

\no where the $\gamma$'s are the conventional gamma matrices. This is the
very reason why the experiment is called the {\it $4\pi$ spinorial
symmetry test}.

In fact, mutations of the intrinsic magnetic moment of nucleons imply a
departure from its characterization via Dirac's equation from which
a departure from law (2.3) follows.  At any rate, for an angle of spin
flip different than $720\c$, spinorial law (2.3) {\sl cannot} represent
the physical setting.

The above occurrences then leave no other choice than the search for a
suitable covering of QM which is more effective for a quantitative
representation of the nuclear aspects under consideration.

In the next section we introduce a new representation of extended,
nonspherical
and deformable hadrons with main characteristics (2.1) and (2.2) which
implies a covering of the spinorial law (2.3) suitable for the exact
representation of $4\pi$ interferometric measures of type (1.3)
and, therefore, of the magnetic moment of few-body nuclei.

Particularly important is the achievement in the next section, apparently
for the first time, of {\sl a generalization of the  spinorial law (2.3)
without altering the spin of nucleons and other QM laws}. The latter advances
are needed to dispel a rather general expectation in the earlier studies
 in the field that a possible confirmation of data (1.3) would imply a
departure from the Fermi-Dirac character of the nucleon, with consequential
inconsistency with established nuclear laws, e.g., Pauli's exclusion
principle. In fact, departures from conventional spin values are present
in the ``spin fluctuations'' of Ref. [4c], the first $SU_{(p,q)}(2)$
quantum group [4e], the proposed test of Pauli's principle under
{\it external strong interactions} [4f] and are inherent in all
subsequent $q$-deformations [4g,4h].

In this paper we present the application of the new formulations of Sect. 3
for the apparently first, exact representation of the total magnetic moments
of few body nuclei. We then show that $4\pi$-interferometric measures can
indeed provide its independent verification. We finally point out other
applications and expected far reaching implications.

\section{Relativistic hadronic mechanics}

The insufficiencies of QM for the nuclear structure as well as for interior
systems in general have been recognized by various research groups. This
has stimulated the appearance in recent decades of various studies on
possible structural generalizations of QM.

A first attempt was that initiated by Santilli [4b] with the {\it parameter}
$(p,q)$-deformations of QM with generalized product
$(A,B)=p\times A\times B-q\times B\times A$ of Albert's Lie-admissible type,
generalized time evolution $i\dot A=(A,H)$, and $SU_{p,q}(2)$ quantum
structure [4e,4f].

By the time Biedenharn [4g] and Macfarlane [4h] initiated their studies of
the simpler class of $(1,q)$-deformations in the mid 1980's (thereafter
followed by a very large number of papers), the author had already abandoned
this line of inquiry because of its rather serious problems of {\sl physical}
consistency identified, e.g., in Refs. [5].

In fact, $(p,q)$-time evolutions are evidently {\it nonunitary}, i.e.,
they have the structure $U\tim U^{\dagger}\neq I\:.$ As such, they
{\sl are not} invariant under their own time evolution which induces the
broader {\it operator} $(P,Q)$-deformations [6a,6b] with generalized
product $(A\hat ,B)=A\tim P\tim B -B\tim Q\tim A$ also of Albert's
Lie-admissible type with $P=q\tim(U\tim U^{\dagger})^{-1}$ and
$Q=q\tim(U\tim U^{\dagger})^{-1}$,
generalized time evolution $i\dot A =(A\hat ,H)$, and correspondingly
broader $SU_{P,Q}(2)$ deformations.

The problematic aspects [5] originate from the fact that, after having been
achieved via nonunitary transforms, the latter structures are themselves not
form-invariant under further nonunitary transforms, thus lacking the
axiomatic consistensy of QM.

More generally, all existing deformations of QM with a nonunitary time
evolution [4], including $q$-, $k$-, quantum-, $(p,q)$- and
$(P,Q)$-deformations, when formulated on {\sl conventional Hilbert spaces
over conventional fields} have the following rather serious problematic
aspects of physical nature [5]: 1) the basic unit is not invariant, thus
preventing unambiguous applications to experiments; 2) Hermiticity is not
conserved in time, thus preventing the existence of unambiguous observables;
3) special functions and transforms are not unique and invariant, thus
implying lack of uniqueness and invariance of
numerical predictions and physical laws; and other problems.

Most importantly, {\sl all the preceding deformations imply the violation of
the special relativity}, e.g., because the deformed Minkowski space and
{\Po} symmetry are not isomorphic to the origin ones (see, e.g., [4j]). This
creates the problem of identifying new axioms replacing Einstein's axioms,
establishing their axiomatic consistency and, after that, proving them
experimentally.

In this paper we shall use a third class of covering formulations [6] which
apparently resolve the above problematic aspects, thus permitting
quantitative studies with invariant basic unit, invariant
Hermiticity-observability,
unique and  invariant special functions, numerical predictions and physical
laws,  yet possessing a nonunitary structure as evidently necessary for
novelty.

Above all, {\sl the covering formulations presented in this section are based
on the central requirement of preserving the axioms of the special relativity
at the abstract level and merely realize them in a more general way}. This
illustrates the reasons for our insistence in using terms different than
``deformations'' [4], such as ``mutations'' [6].

The emerging theory is known under the name of {\sl hadronic mechanics} (HM),
today also known (for certain reasons identified below) as {\sl ``isotopic
completion'' of quantum mechanics}, as originally proposed by the author in
[6a,6b] and subsequently studied by numerous researchers (see [7] for
independent studies and comprehensive bibliographies), and outlined in the
recent monographs [6j,6k].

The formulation which is necessary for the study of the historical hypothesis
of Fig. 1, and its application to total nuclear magnetic moments is
{\sl relativistic hadronic mechanics} (RHM) or {\sl ``isotopic completion''}
of RQM. Its study in Refs. [6j,6k] is made for the most general possible
mutations as apparently needed for extreme, interior, hadronic and
astrophysical conditions.

In this section we shall present the apparently first formulation of RHM
specifically conceived for nuclear physics under the crucial condition of
representing the historical hypothesis of Fig. 1. In so doing, we shall
show, also for the first time, that RHM can provide the above representation
while preserving all conventional QM laws, such as Heisenberg's
uncertainties, Pauli's exclusion principle, etc.

RHM is constructed via maps of RQM called {\it isotopies} [6a] from the
Greek meaning of being ``axiom-preserving" and referred to {\sl maps of any
given linear, local-differential and unitary theory into its most general
possible nonlinear, nonlocal-integral and nonunitary extensions which are
nevertheless capable of reconstructing linearity, locality and unitarity in
certain generalized spaces called {\it isospaces}, and generalized fields
called {\it isofields}}.

It then follows that isotopic images of fields, spaces, algebras, etc., are
isomorphic to the original structures by conception and construction, and they
coincide at the abstract, realization free level, all this as preparatory
grounds to preserve Einsteinian axioms of the special relativity.
Nevertheless, as we shall see shortly, the two theories are physically
inequivalent because connected by {\sl nonunitary transforms}.

Recall that the most dominant aspect of the predicted new terms in the nuclear
force is that of {\sl not being representable with a Hamiltonian} and,
of being {\sl nonunitary} (otherwise we trivially remain within the class
of equivalence of RQM). The best way to construct the foundations of RHM is
therefore by subjecting the corresponding foundations of RQM to nonunitary
transforms.

The fundamental quantities of RQM are: the basic unit of the underlying
Minkowski space, $I=\diag\,(\{1,1,1\},1)$ in Euclidean space (say, 1 cm)
and the unit of time (say, 1 sec) in dimensionless form with $\hbar=1$;
the basic associative product $A\tim B$ among generic quantities $A$, $B$
(which is the same for all products of RQM, those of: numbers, operators,
etc., including the modular action $H\tim |\psi\rangle$ of operators $H$ on
Hilbert states $|\psi\rangle$); and the fundamental relativistic canonical
commutation rules $[p_{\mu},x^{\nu}]=
p_{\mu}\tim x^{\nu}-x^{\nu}\tim p_{\mu}=-i{\delta_{\mu}}^{\nu}\tim I\:.$

Under nonunitary transforms, the above quantities become

\def\h{\hat}
\def\arr{\rightarrow}
\def\dag{\dagger}

$$
U\tim U^{\dagger}=\hat I={\h I}^{\dagger}\neq I\:,\eqno{(3.1a)}
$$
$$
I\arr \h I=U\tim I\tim U^{\dagger}\:,\eqno{(3.1b)}
$$
$$
A\tim B\arr\h A\h{\tim}\h B=U\tim A\tim B\tim U^{\dagger}=
\h A\tim\h T\tim\h B\:,\eqno{(3.1c)}
$$
$$
U\tim [p_{\mu},x^{\nu}]\tim U^{\dag}=[{\h p}_{\mu}\h ,{\h x}^{\nu}]=
{\h p}_{\mu}\h{\tim} x^{\nu} -{\h x}^{\nu}\h{\tim}{\h p}_{\mu}=
-i{\delta_{\mu}}^{\nu}\tim\h I\:,\eqno{(3.1d)}
$$
$$
\h T=(U\tim U^{\dag})^{-1}=\h I^{-1}\:,\quad\h K=U\tim K\tim U^{\dag}\:,
\quad K=A,B, p, x\:.\eqno{(3.1e)}
$$

\no The above new images are then assumed as the fundamental quantities of
RHM.

A most dominant aspect of the above nonunitary transforms is that they
imply the joint mapping, called {\it lifting} [6a], of the unit $I\arr\h I$
while the product is lifted in an amount which is the {\it inverse} of that
of the unit, $A\tim B\arr\h A\h{\tim}\h B=\h A\tim \h T\tim \h B$, under
which $\h I={\h T}^{-1}$ is the correct left and right unit of the new
theory,

$$
\h I\h{\tim}\h A={\h T}^{-1}\tim\h T\tim\h A\equiv \h A\h{\tim}\h I=
\h A\tim\h T\tim{\h T}^{-1}\equiv A\:,\quad \forall A\:,\eqno{(3.2)}
$$

\no which case (only) $\h I$ is called the {\it isounit} and $\h T$ the
{\it isotopic element} [6a,6b].

The emerging new operator envelope $\xi$ is called {\it isoassociative}
because it verifies the associative law with respect to the isoproduct,
$\h A\h{\tim}(\h B \tim\h C)=(\h A\h{\tim}\h B)\h{\tim}\h C$. Note that the
new unit $\h I$ is Hermitean and will therefore be assumed hereon to be
positive-definite. Under these conditions it is evident that {\sl the original
envelope $\xi$ and its isotopic image $\h{\xi}$ are isomorphic by central
objective, $\xi\approx\h{\xi}$, and the map $\xi\arr\h{\xi}$ is an
{\it isotopy}. Yet they are physically nonequivalent because nonunitarily
related.}

The representation of system with RQM is done via the knowledge of
{\sl one operator only}, the Hamiltonian $H$, {\sl under the tacit
assumption of the simplest possible basic units} $I=\diag\,(\{1,1,1\},1)\:.$
The representation of systems via RHM requires the knowledge of {\sl two}
quantities, the conventional Hamiltonian $H$ to represent conventional
potential interactions, and a second quantity, the isounit $\h I$, to
represent all nonhamiltonian quantities.

We shall therefore assume hereon the realization of the isounit
(for $\hbar =1$),

$$
\h I=\diag\,({\h I}_s,{\h I}_t)=
\diag\,(\{{n_1}^2,{n_2}^2,{n_3}^2\},
{n_4}^2)\tim\h{\Gamma}(x,\dot x,\psi,\partial\psi ,\ldots )> 0\:,\eqno{(3.3a)}
$$
$$
{\h I}_s=\diag\,\{{n_1}^2,{n_2}^2,{n_3}^2\}\tim\h{\Gamma}_s
(x,\dot x,\psi ,\partial\psi ,\ldots)\:,
{\hat I}_t={n_4}^2\tim\h{\Gamma}_t
(x,\dot x,\psi,\partial\psi ,\ldots)\:,\eqno{(3.3b)}
$$

\no where $\h I_s$ and $\h I_t$ are called the {\it space and time isounits},
respectively, the ${n_k}^2$'s are quantities (2.1) representing the shape
of the hadron considered, ${n_4}^2$ is the quantity (2.2) representing the
density of the medium in which motion occurs, and $\h{\Gamma}$ is a
positive-definite $4\tim 4$ matrix representing the contact, nonlinear,
nonlocal, nonhamiltonian and nonunitary interactions (as identified in
Sect. 2). The functional dependence of the isounit remains completely
unrestricted in RHM and must be determined from the characteristics of the
case at hand exactly as it is the case of the Hamiltonian in RQM.

It is important to know that identifications (3.3) are made following the
historical teaching by  Hamilton, Lagrange, Jacobi and other Founders of
analytic dynamics according to which one quantity alone (we today call
the Hamiltonian or the Lagrangian) simply cannot represent the entire physical
reality. For this reason they formulated their analytic equations with
{\it external terms}.

Comprehensive classical studies not reported here for brevity (see the
review and bibliography in [6j,6k]) have established that the use  of isounits
(3.3) is analytically equivalent to external terms and, in fact, they have
the same number of independent elements.  The reformulation of the external
terms via the isounit has resulted to be necessary to preserve Einsteinian
axioms of the special relativity beginning at the classical level because
it permits the preservation of Lie's theory which would be otherwise
lost in favor of the broader Lie-admissible theory [6a].

These classical studies have resulted in a new analytic mechanics, called
{\it isohamiltonian mechanics} which is the unique and unambiguous classical
image of the operator mechanics outlined in this section (see [6l] for the
latest studies and comprehensive bibliography).

A criticism is at times moved according to which RHM is ``too broad'' because
the isounit can have infinitely possible values. This criticism is evidently
equivalent to the statement that RQM is ``too broad" because it admits
infinitely possible Hamiltonians.

In reality, both the Hamiltonian and the isounits are selected via fully
identified methods resulting in all applications considered until now
in unique and unambiguous expressions. The Hamiltonian is selected via all
conventional criteria which are those of the {\it exterior problems}, such as
mass, charge, potential, etc. The isounit is instead selected on the {\sl new}
grounds of the {\it interior problems}, thus requiring the description
of extended, nonspherical and deformable shapes, density/index of refraction
and contact interactions which are absent in the QM literature of this
century.

At any rate, any quantitative representation of the historical hypothesis
of  Fig. 1 requires the capability to represent all {\sl infinitely possible
different shapes of the same nucleon}, thus requiring for consistency
infinitely many possible isounits for {\sl each} given Hamiltonian.

Also, the noninitiated reader should know since these introductory lines that,
when an isolated interior system is considered from the outside, {\sl internal
nonpotential effects must evidently averaged into constants because they
are short range}, by therefore resulting in a mere rescaling of the shape and
density terms via the constant factor $\hat{\Gamma}_0=
\langle\hat{\Gamma}\rangle\:.$ This
occurrence renders preferable the scale invariant description of the
characteristic $n$-quantities, which will be tacitly adopted hereon.

Once the basic isotopic unit, product and commutation rules are known, the
next step is the identification of the axiomatically correct structure of RHM.
Extensive studies in this respect completed only recently with the appearance
of Ref. [6l] have shown that {\sl RHM is as axiomatically consistent as RQM
if and only if the nonunitary maps (3.1) are applied to the totality of the
formalism of RQM, without any exception known to the author}. In fact, any
mixtures of isotopic structure with conventional QM methods leads to a host of
inconsistencies which generally remain undetected by nonexperts in the field.

This implies that the formalism of RQM must be reconstructed in such a way
to admit $\hat I$, rather than 1, as the correct left and right unit. Thus,
numbers, metric spaces, geometries, symmetries, Hilbert spaces, etc., have to
be reconstructed in terms of the isoproduct $A\h{\tim}B$ with isounit $\h I$.
The construction is simple, yet unique and unambiguous, and is done below for
the first time by deriving each new structure from the single nonunitary map
(3.1), under the notation according to which all quantities with a ``hat'' are
computed in generalized spaces and those without are computed in conventional
spaces.


\subsection{Isofields}

The first notion of RQM which must be isotopically
lifted in order to achieve invariant units, Hermiticity and numerical
predictions is that of the fields of ordinary real numbers
${\cal R}(n,+,\tim)$ and complex numbers ${\cal C}(c,+,\tim)$ with
conventional sum $a+b$, additive unit 0, multiplication $a\tim b$ and
multiplicative unit $I$, $a=n$, $c$, resulting
in the {\it isofields} [6f] $\h{{\cal R}}=\h{\cal R}(\h n,+,\h{\tim})$ and
$\h{\cc}(\h c,+,\h{\tim})$ of {\it isoreal numbers}
$\h n=U\tim n\tim U^{\dag}=
n\tim\h I$ and {\it isocomplex numbers} $\h c=U\tim c\tim U^{\dag}=
c\tim \h I$, $n\in{\cal R}$, $c\in \cc$, $\h I\neq{\cal R},\cc$ equipped with
the conventional sum $\h +\equiv +$ and related additive unit $\h 0\equiv 0$,
as well as with the isoproduct and related isounit

$$
\h a\h{\tim}\h b=U\tim a\tim b\tim U^{\dag}=\h a\tim\h T\tim\h b\:,\quad
\h I={\h T}^{-1}\:,
$$
$$
\h I\h{\tim}\h a\equiv\h a\h{\tim}\h I\equiv\h a\:,
\:\:\forall a=n, c\:.\eqno{(3.4)}
$$

\no The important property is that $\h{\cal R}$ and $\cc $ preserve all axioms
of a field [6j]. Thus, the liftings ${\cal R}\arr\h{\cal R}$ and
$\cc\arr\h{\cc }$ are isotopies.

For consistency, all operations on numbers must then be isotopically lifted in
a simple yet unique and significant way. We have in this way the following
{\it isosquare, isosquare root, isoquotient, isonorm,} etc. (see [6f] for
details).

$$
\h a^2=\h a\h{\tim }\h a=a^2\tim\h I\:,\quad\h a^{\h{1\over 2}}=a^{1\over 1}
\tim\h I^{1\over 2}\:,
$$
$$
\h a\h /\h b=(\h a/\h b)\tim \h I\:,\quad
\h |\h a|=|a|\tim\h I\:,\quad a=n,c\:.\eqno{(3.5)}
$$

Thus the tradition statement ``$2\tim 2=4$'' remained unchanged since
biblical times has meaning for RQM but has no meaning for RHM because one
must identify first the selected unit and product for the operation
``$2\tim 2$'' to have sense. This illustrates from the outset the insidious
inconsistencies in attempting to appraise the new RHM via the use of old
mathematics.

In short, RQM is defined for numbers $n$ whose basic unit is the quantity
$+1$ dating back to biblical times. RHM is instead defined for new numbers
$\h n=n\tim\h I$ which admit arbitrary (positive-definite) units $\h I$. As
we shall see shortly, the introduction of the new isonumbers has deep and
intriguing implications, including the possibility of defining {\sl new}
symmetries for {\sl conventional} line elements and inner products.


\subsection{Iso-Hilbert spaces}

The second notion of RQM which must be lifted
for consistency is that of conventional Hilbert spaces ${\cal H}$ with states
$|\psi\rangle ,|\phi\rangle ,\ldots ,$ inner product
$\langle\phi |\psi\rangle\in\cc $ and
normalization $\langle\psi |\psi\rangle =1$, resulting in the
{\it iso-Hilbert space}
$\h{\cal H}$ [6j] with the following {\it isostates, isoinner product} and
{\it isonormalization}

$$
\langle\h{\phi}\h |\h{\psi}\rangle =
U\tim\langle\phi |\tim U^{\dag}\tim(U\tim U^{\dag})^{-1}\tim
U\tim |\psi\rangle\tim U^{\dag}=
$$
$$
=\langle\h{\phi}|\tim\h T\tim |\h{\psi}\rangle\tim\h I\in
\h{\mbox{\bf C}}\:,\eqno{(3.6a)}
$$

$$
\langle\h{\psi}|\tim\h T\tim |\h{\psi}\rangle =1\:,\quad |\h{\psi}\rangle =
U\tim |\h{\psi}\rangle\:,
\quad \langle\h{\phi}|=\langle\phi |\tim U^{\dag}\:.\eqno{(3.6b)}
$$

Note that, again for consistency, the isoinner product must be an isocomplex
number, i.e., must have the structure $\h c=c\tim\h I$. The new composition
is still inner (because $\h T\r 0$) and, therefore, $\h{\cal H}$ is still
Hilbert. Then, $\h{\cal H}\approx{\cal H}$ and the lifting ${\cal H}\arr
\h{\cal H}$ is again an isotopy.

The local isomorphism ${\cal H}\approx\h{\cal H}$ can also be seen from the
following {\sl new invariance law of the conventional Hilbert product} here
expressed for $\h T$ independent from the integration variable,

$$
\l \phi |\psi \r \tim I\equiv \l \phi |\tim |\psi \r \tim\h T\tim\h T^{-1}
\equiv\l \phi |\tim\h T\tim |\psi \r \tim\h I\equiv\l \phi\h |\psi
\r \:.\eqno{(3.7)}
$$

\no Thus, RHM is based on conventional  Hilbert spaces, only realized in a way
more general than that of current use.
8
\def\hh{\hat{\times}}
\def\hps{\hat{\psi}}
\def\hph{\hat{\phi}}

The isotopy ${\cal H}\arr\h{\cal H}$ is equally fundamental for the
consistency
of the theory. To see it, note that, under the lifting $I\arr\h I=\h T^{-1}$
and $A\tim B\arr\h A\h{\tim}\h B=\h A\hh\h T\hh \h B$, the action of an
operator
$H$ on a state must be isotopic, i.e., of the type $\h H\hh |\hps\r =\h H\tim
\h T\tim |\hps \r $ because this is the only one admitting the isounit
$\h I\hh |\hps \r \equiv |\hps \r $. Then  the formulation of the above
expression
on a {\sl conventional} Hilbert space with inner product
$\l \hph |\tim |\hps \r $
implies the general  loss of Hermiticity.  In fact, we would have the
condition
$\{\l \hph |\hh \h H^{\h{\dag}}\}\tim |\hps \r =\l \hph |\tim\{\h H\hh |
\hps \r \}$, i.e., $\h H^{\h{\dag}}=
\h T^{-1}\tim H^{\dag}\tim \h T\neq \h H^{\dag}$.
On the contrary, the use of the isoHilbert space implies the conditions

$$
\{\l \hph |\hh\h H^{\h{\dag}}\}\hh |\hps \r =
\l \hph |\hh\{\h H\hh |\hps \r \}\:,
\quad \mbox{i.e.}\:,\quad \h H^{\h{\dag}}=\h H^{\dag}\:.\eqno{(3.8)}
$$

As a result, {\sl the conditions of Hermiticity and  isohermiticity coincide,
quantities which are Hermitean-observable for RQM remain so for RHM, the
eigenvalues of Hermitean operators  of RHM are real}, and other properties
(see [6j] from brevity).

The only possible {\sl isoeigenvalues equations} are then given by

$$
\h H\hh |\hps \r =\h H(x,p)\tim \h T(x,p,\psi ,\partial\psi ,\ldots )\tim
|\hps\r =\h E\hh |\hps \r =
$$
$$
=(E\tim \h I)\tim\h T\tim |\hps \r =E\tim |\hps \r \:.\eqno{(3.9)}
$$

\no with corresponding {\sl isotopic expectation values}

$$
\l \h H\r =\frac{\l \hps |\hh\h H\hh |\hps \r }{\l \hps |\hh |\hps\r }=
\frac{\l \hps |\tim\h T\tim\h H\tim\h T\tim |\hps \r }{\l \hps |\tim\h T\tim |
\hps \r }\:,\eqno{(3.10)}
$$

\no which can be easily seen to coincide with the  isoeigenvalues for the same
operator. Note from Eq.s (3.9) that the ``final numbers" of RHM to be
confronted with experiments are conventional.

The fundamental axioms of RHM, are a simple isotopy of the axioms of RQM here
omitted for brevity [6k]. We only mention for future needs the axiom.

The above elements illustrate the main property that {\sl RHM coincides with
RQM at the abstract realization-free level} for which, from the
positive-definiteness of $\h I$, we have  ${\cal R}\equiv\h{\cal R}$,
$\mbox{\bf C}\equiv\h{\mbox{\bf C}}$ and ${\cal H}\equiv\h{\cal H}$. All other
aspects of RHM are constructed following the same lines. Thus, {\sl RHM is not
a new theory, but merely a new realization of the abstract axioms of RHM}.
These  properties then establish the axiomatic consistency of RHM to such an
extent that any criticism in its axiomatic structure is  {\sl de facto} a
criticism on the axiomatic structure of RQM.

Despite the above abstract axiomatic identity, one should keep in mind that,
as illustrated  in Eq.s (3.1), {\sl RHM and  RQM are physically inequivalent
because the former is a nonunitary image of the latter}. Moreover, isotopies
imply the following mapping of eigenvalues

$$
H\tim |\psi \r =E_0\tim |\psi \r \arr H\tim T\tim |\psi \r =
E\tim |\hps \r \:,\quad E\neq E_0\:,\eqno{(3.11)}
$$

\no according to which  {\sl the same operator $H$ has different eigenvalues
in RQM and  RHM}, and this illustrates the nontriviality of the isotopies.


\subsection{Isolinearity, isolocality, isounitarity}

It is important to see that, despite their physical inequivalence, {\sl RHM
preserves the conventional linearity, locality and unitarity of RQM}.
To begin,
RHM is highly nonlinear in the weavefunctions (and their derivatives), as
evident from isoeigenvalues expressions (3.9). Yet, the theory is {\sl
isolinear}, i.e., it verifies the linearity conditions in isospace, e.g., for
all possible $\h a\in \h{\cal R}$ or $\h{\cc}$ and
$|\hph \r ,|\hps \r \in\h{\cal H}$, we have the identity

$$
\h A\hh(\h a\hh |\hps \r +\h b\hh |\hph \r )=\h a\hh\h A\hh |\hps \r +
\h b\hh\h A\hh |\hph \r \:,\eqno{(3.12)}
$$

A similar situation occurs for {\sl locality}. In fact, RHM is {\sl
nonlocal-integral} because interactions of that type are admitted in the
$\h{\Gamma}$'s terms of the isounits, Eq.s (3.3). Nevertheless, RHM is
{\sl isolocal}, i.e., it verifies the condition of locality in isospace. In
particular, RHM is everywhere local-differential except at the isounit.
On more technical grounds, RHM is equipped with a new topology called
{\sl Tsagas-Sourlas integro-differential topology} [7f].

By recalling that RQM is strictly local-differential, the above new topology
has fundamental physical relevance inasmuch as it permits mathematically
rigorous quantitative studies of the nonlocal-integral component of the
nuclear force needed to represent the overlapping of the hyperdense charge
distributions of nucleons in the nuclear structure [5f].

Next, RQM is said to be  {\sl unitary} in the sense that the only allowed
transformations are of the unitary type, $U\tim U^{\dag}=U^{\dag}\tim U=I$.
By comparison, RHM is {\sl nonunitary} because its transformation theory is
based on the requirement  $W\tim W^{\dag}=\h I\neq I$. Nevertheless, RHM
reconstructs unitarity in isospace, a property called {\sl isounitarity}. In
fact, the above nonunitary transforms cal be rewritten in the following
{\sl identical} isotopic form

$$
W=\h W\tim\h T^{1/2}\:,\quad W\tim W^{\dag}=\h I\neq I\:,\eqno{(3.13a)}
$$

$$
W\tim W^{\dag}\equiv\h W\hh\h W^{\dag}=
\h W^{\dag}\hh\h W=\h I\:,\eqno{(3.13b)}
$$

\def\e#1{\eqno{(#1)}}

The necessity of the preceding reformulation is soon established by the fact
that,even though derived via nonunitarity transforms, the isotopic
structures (3.1) and related properties {\sl are not} invariant under
additional nonunitary transforms. However, the needed form-invariance is
readily achieved under the isounitary reformulation (3.13) for which

$$
\h I\arr\h I=\h W\hh\h I\hh\h W^{\dag}=\h W\tim\h T\tim\h T^{-1}\tim \h T\tim
\h W^{\dag}\equiv\h I\:,\e{3.14a}
$$

$$
\h A\hh\h B\arr\h W\hh\h A\hh\h B\hh\h W^{\dag}=\h A'\hh\h B'\:,
$$
$$
\h K'=\h W\hh\h K\hh\h W^{\dag}\:,\quad\h K v=\h A < \h B\:,\e{3.14b}
$$

\no with a corresponding invariance of the condition of isohermiticity and all
other properties [6k]. This illustrates again that the lack of application of
the isotopies to {\sl any} aspect of RQM implies insidious axiomatic
inconsistencies.

Note that under isotransforms (3.14) the isounit and isotopic element remain
{\sl numerically invariant}. Note also that the transformation theory of RQM
is restricted to transforms verifying the condition $U\tim U^{\dag}=I$ for
a fixed $I$. Similarly, the isotransforms of RHM are restricted to those
verifying the condition $\h W\hh\h W^{\dag}=\h I$,
this time, for fixed $\h I$ (because its change would imply the description
of a {\sl different} system).


\subsection{Isotopic physical laws}

In this paper we are presenting the simplest possible branch of RHM, that
specifically  formulated for applications  to nuclear physics via a {\sl
diagonal, Hermitean and positive} isounit (3.2). It is easy to see that the
above branch does indeed preserve all conventional QM laws.

Recall that {\sl generalizations of RQM which are conventionally nonlinear
in the wavefunctions, i.e. of the type} $H(x, p,\psi ,\ldots)\tim\psi=
E\tim\psi$ [8] {\sl imply the loss of the superposition principle, with
consequential inapplicability to a consistent treatment of composite systems
such as nuclei}, besides having additional problematic aspects studied in
[6k,9].

RHM is also highly nonlinear in view of the eigenvalue structure (3.9), i.e.,
$H(x,p)\tim\h T(\hps ,\ldots)\tim\hps=E\tim\hps$. However, the mathematical
notion of isolinearity has the important physical implication that {\sl RHM
preserves the superposition principle in isospace}, as one can verify. This
has the important implication that RHM can indeed be consistently applied to
composite systems such as  few-body nuclei.

Moreover, conventional nonlinear systems can be  {\sl identically}
reformulated
in the isotopic form, $H(t,r,\psi)\tim\psi\equiv H_0(t,r)\tim\h T(\psi ,\ldots
)\r \psi =E\tim\psi$, by therefore recovering axiomatic consistency in
isospace.

Next, it is important to see that {\sl RHM preserves  Heisenberg's uncertainty
principle}. In fact, from isocommutators  (3.1d) we  have ($\hbar =1$)

$$
\Delta\h r^i\Delta\h p_j\geq{1\over 2}\l [\h r^i\h ,\h p_j]\r ={1\over 2}
{\delta^i}_j\:.\e{3.15}
$$

\no This  establishes that the deviations from Heisenberg's uncertainties
predicted by quantum deformations (e.g., of  the so-called squeezed states
[4j]) can be removed via their reformulation in an invariant isotopic form
(see [6k] for details).

Along similar lines, it is possible to prove that the notions of isolocality
and isounitarity permit the preservation of {\sl causality under
nonlocal-integral forces} (see also [6k] for brevity). The preservation of the
Fermi-Dirac statistics and related Pauli's exlusion principle will be
indicated shortly. The proof of the preservation of other physical laws will
be left to the interested reader.

The preservation of conventional laws can be seen from the fact that the
fundamental quantity of representing deviations from RQM, the isounit,
preserves all axiomatic properties of the  conventional unit $I$, it is the
basic invariant of the new theory and its isoexpectation values recover the
conventional value $I$,

$$
\h I^{\h n}=\h I\hh\h I\hh\ldots\hh\h I\equiv\h I\:,\quad \h I^{\h{1\over 2}}
=\h I\:,\quad \h I\h /\h I\equiv \h I\:,\quad\mbox{etc.}\e{3.16a}
$$

$$
\h I'=\h W\hh\h I\hh\h W^{\dag}\equiv\h I\;,\quad id\h I/dt=\h I\hh\h H-\h H
\hh\h O=\h H-\h H\equiv 0\:,\e{3.16b}
$$

\def\hi{\hat I}

$$
\l \h I\r =\frac{\l \hps |\hh\hi\hh |\hps \r }{\l \hps |\hh |\hps \r }=
\frac{\l \hps |\tim\h T\tim
\h T^{-1}\tim \h T\tim |\hps \r }{\hps |\tim\h T\tim |\hps \r }=I\e{3.16c}
$$

The above properties establishes the occurrence with far reaching implications
according to which {\sl the validity of conventional QM laws for the nuclear
structure, such as Heisenberg's uncertainty principle, Pauli's exclusion
principle, etc., by no means, imply that the conventional formulation of
RQM is the  only applicable discipline because exactly the same laws are
admitted  by the structurally more general RHM.}

It should be indicated for completeness that in this paper we are studying the
simplest possible realization of RHM, that specifically constructed for the
nuclear structure under the condition of preserving conventional physical
laws. More general realizations exist [6k,6l], e.g., those still of isotopic
type with {\sl nondiagonal} isounit $\h I$, or the more general ones of
{\sl genotopic type} with nonhermitean basic unit and related transforms

$$
I\arr\h I=U\tim I\tim W^{\dag}\neq\h I^{\dag}\:,\quad U\tim U\neq I\:,\quad
W\tim W^{\dag}\neq I\:,\e{3.17}
$$

\no which are particularly suited to represent {\sl irreversibility under
open-noncon\-servative conditions}, or those of {\sl hyperstructural} type
where $\h I$
is a {\sl set} of nonhermitean elements, which are particularly suited to
represent
irreversible biological systems [6k], for which conventional QM laws are not
necessarily preserved.

The formulation of RHM presented in this section is intended to describe
nucleons
when members of a nuclear structure with {\sl conventional} spin verifying
{\sl conventional} laws and merely having  a {\sl nonspherical-deformable}
shape.
The more general formulations indicated above are intended for more general
physical conditions, such as a neutron in the core of a collapsing star
considered
as external, and will not be considered in this paper.


\subsection{Isotopic realization of ``hidden variables" \hfil\break and
EPR ``completion" of RQM}

The reader should be aware that {\sl RHM provides an explicit and
concrete realization
of the theory of ``hidden variables"} [10a], which are actually
realized via the
{\sl operator} $\lambda=\h T(x,\dot x,\hat{\:},\partial\psi ,\ldots)$ and
isoeigenvalues

$$
\h H\hh_{\lambda}|\hps \r =\h H\tim\lambda
(x,\dot x ,\psi ,\partial\psi ,\ldots)\tim |
\hps \r = E_{\lambda}\tim |\hps \r \:.\e{3.18}
$$

\no In fact, the right modular actions ``$H\tim\psi$" and ``$H\hh\psi$" lose
any distinction at the abstract level and, in this sense, they are evidently
``hidden" in the conventional realization.

As a result, {\sl RHM constitutes a form of ``completion" of RQM, hereon
called "isotopic completion", which results to be much along the celebrated
argument by  Einstein, Podolsky and Rosen } [10b]. In particular, the
completion is permitted by the fact that von Neumann's theorem [10c] and
Bell's inequalities [10d] are {\sl inapplicable } (and not "violated") for
the isotopic completion due to its {\sl nonunitary} structure.

More specifically, von Neumann theorem is inapplicable because the {\sl same}
Hamiltonian $H$ has an infinite number of {\sl different} sets of eigenvalues
in RHM, one per each possible isotopic element ("hidden operator")
$\l =\h T$. Bell's inequalities are inapplicable, e.g., because RHM
requires a {\sl nonunitary image of Pauli's matrices} (see below for their
outline). For detailed studies  see ref. [6k], App. 4.C. The classical limit
under isotopies is also studies in ref. [10e].

A consequence of the above occurrences  is that {\sl all applications of RHM
outlined below, including the exact representation of nuclear magnetic moments
of Sect. 4, are applications of the  "isotopic completion" of RQM much along
the celebrated $E=P-R$ argument}.


\subsection{Iso-Minkowski spaces}

The next notion of RQM which must be isotopically lifted for compatibility
with basic structures (3.1) is that of the underlying carrier space, the
Minkowskian space $M(x,\eta,{\cal R})$ with space-time coordinates
$x=\{x^{\mu}\}=\{r,c_0t\}$, where $c_0$ is the speed of light in vacuum, and
metric $\eta=\diag\,(1,1,1,\\-1)$ and basic unit $I=\diag\,
(\{1,1,1\},1)$ on $\cal R$. The listing yields the {\it iso-Minkowski space}
$\h M=\h M(\h x,\h{\eta},\h{\cal R})$, first proposed by Santilli [6e] in
1983, which is characterized by the lifting of: the coordinates $x$ into the
{\it isocoordinates} $\h x=U\tim x\tim U^{\dag}=x\tim\h I$; the basic unit of
$M$ into the isounit (here assumed to be diagonal from its Hermiticity),
$I\arr\h I$, and the lifting of the metric $\eta$ of the {\sl inverse} of that
of the unit, $\eta\arr\h{\eta}=\h T\tim\eta$. The basic {\it isointerval}
is in then given by or $\h x,\h y\in\h M$

\def\parr{\partial}

$$
(\h x-\h y)^{\h 2}=[(\h x^{\mu}-\h y^{\nu})\hh\h N_{\mu\nu}(x,\dot x,
\hps ,\parr\psi ,\ldots)\hh(\h x^{\nu}-\h y^{\nu})]\hh\h I=
$$
$$
=(x-y)^{\h 2}=[(x^{\mu}-y^{\nu})\tim\h{\eta}_{\mu\nu}(x,\dot x,\hps ,
\parr\psi ,\ldots)\tim(x^{\nu}-y^{\nu})]\tim\h I=
$$
$$
=[(x_1-y_1)^2{\h T_1}^{\:\: 1}+(x_2-y_2)^2{\h T_2}^{\:\: 2}-(x_3-y_3)^2
{\h T_3}^{\:\: 3}-
(x_4-y_4)^2{\h T_4}^{\:\:4}]\tim\h I\in\h{\cal R}\:.\e{3.19a}
$$

$$
\h I=\diag\,(\{ {\h I}^1_{\:\:1},{\h I^2}_{\:\: 2},
{\h I^3}_{\:\: 3}\},{\h I^4}_{\:\: 4})=\h T^{-1}> 0\:,
$$
$$
\h T=\diag\,(\{{\h T_1}^{\:\:1},{\h T_2}^{\:\: 2},
{\h T_3}^{\:\: 3}\},
{\h T_4}^{\:\: 4})> 0\:,\e{3.19b}
$$
\no where $\h N$ is an {\it isomatrix}, i.e., a matrix whose elements are
isoscalars $\h N_{\mu\nu}=\h \eta_{\mu\nu}\tim\h I\in\h {\cal R}$ (and,
therefore, its operations and products are isotopic) while  $\h{\eta}$ is an
ordinary matrix, i.e., with elements $\h{\eta}_{\mu\nu}$ given by ordinary
scalars.

Note from the preceding structure that the use of isocoordinates $\h x=x\tim
\h I$ is redundant in the isointerval. Nevertheless we shall keep using the
scripture $\h M=\h M(\h x,\h{\eta},\h{\cal R})$ rather than $\h M(x,\h{\eta},
\h{\cal R})$ to recall that the coordinates are computed in isospace with
respect to a generalized metric. Note also that the isospace $\h M$ and the
underlying isofield $\h{\cal R}$ share the same generalized unit $\h I$.

Note finally that $\h M$ constitutes the most general possible invariant
with signature $(+,+,+,-)$ with a well behaved, yet arbitrary functional
dependence on coordinates, wavefunctions, their derivatives of the needed
order, as well as any additional quantity of the interior problem.

For this reason, as shown  in details by Aringazin [11], the iso-Minkowski
space $\h M(\h x,\h{\eta},\h{\cal R})$ is said to be {\sl directly universal},
i.e., admitting as particular case all possible signature-preserving
generalizations of $M$ (universality), directly in the coordinates of the
observer (direct universality). In particular, the iso-Minkowskian metric
$\h{\eta}$ admits as particular cases the Riemannian, Finslerian,
non-Desarguesian and all other possible metrics in (3+1)-dimension.

Despite the above arbitrariness, it has been proved that the original
(abstract) Minkowskian axioms are preserved under the joint liftings
$I\arr\h I=\h T^{-1}$ and $\eta\arr\h{\eta}=\h T\tim\eta$. Thus,
$\h M\approx M$ and the lifting $M\arr\h M$ is an isotopy.

The latter results can also be seen via the  {\sl new invariance law of the
conventional Minkowskian interval} (here expressed for a non-null scalar
function $n$)

\def\parr{\partial}

$$
(x-y)^2=[(x^{\mu}-y^{\mu})\tim\eta_{\mu\nu}\tim(x^{\nu}-y^{\nu})]\tim I\equiv
$$
$$
\equiv [(x^{\mu}-y^{\mu})\tim(n^{-2}\tim\eta_{\mu\nu})\tim(x^{\nu}-y^{\nu})]
\tim(n^2\tim I)=
$$
$$
=[(x^{\mu}-y^{\mu})\tim\h{\eta}_{\mu\nu}\tim(x^{\nu}-y^{\nu})]\tim \h I=
(x-y)^{\h 2}\:,\e{3.20}
$$

\def\mb{\mbox}

\no The new invariance identified by RHM is therefore $[L=\mb{length}]\tim
[I=\mb{unit}]=\mb{Inv}$. As we shall see shortly, this is the mechanism
which permits the preservation of spin and other conventinal laws.

Thus, RHM merely expresses "hidden" degrees of freedom of conventional
quantum axioms. These degrees of freedom, expressed via the new invariance
laws (3.7) and (3.20) have  remained undetected through this century because
they required the prior discovery of {\sl new numbers}, those with arbitrary
units [6f].


\subsection{Isodifferential calculus}

Despite the use of the isotransformations
theory, dynamical equations on $\h M$ {\sl are not} invariant when expressed
in terms of the conventional differential calculus. This has requested the
construction of the {\sl isodifferential calculus} [6l] which is characterized
by a simple, yet unique and effective isotopy of the conventional calculus
based on the following {\sl isodifferential, isoderivative} and related
primary properties

$$
\h d\h x^{\mu}={\h I^{\mu}}_{\:\:\alpha}\tim d\h x^{\alpha}\;,\quad
\h{\parr} /\h{\parr}\h x^{\mu}={\h T_{\mu}}^{\:\:\alpha}\tim\parr /\parr\h
x^{\alpha}\:,\e{3.21a}
$$

$$
\h{\parr}\h x^{\mu}\h /\h{\parr}\h x^{\nu}={\delta^{\mu}}_{\nu}\tim\h I\:,
\quad
\h{\parr}\h x^{\mu}\h /\h{\parr}\h x_{\nu}=\h I^{\mu\nu}\tim\h I\:,\quad
\h{\parr}\h x_{\mu}\h /\h{\parr}\h x^{\nu}=\h T_{\mu\nu}\tim\h I\:,
\e{3.21b}
$$

\no where we have implied identities of the type

$$
\h x\hh|\hps \r =x\tim |\hps \r \:,\quad
(\h{\parr}\h /\h{\parr}\h x)\hh |\hps \r \equiv\h{\parr}/\h{\parr}\h x|\hps
\r \:,\quad\mb{etc.}\e{3.22}
$$

The above isocalculus has  only recently permitted the achievement of an
axiomatically consistent and form-invariant characterization of the
{\sl isotopic linear momentum operator} [6l] which had escaped identification
for over a decade and which can be written ($\hbar =1$)

$$
\h p_{\mu}\hh |\hps \r =p_{\mu}\tim\h T(x,\dot x,\psi ,\parr\psi ,\ldots)\tim
|\hps \r =-i\h{\parr}_{\mu}|\hps \r =-i{\h T_{\mu}}^{\:\:\alpha}\parr_{\alpha}
|\hps\r \:,\e{3.23}
$$

\no which does indeed recover the fundamental isocommutation rules (3.1d).

Isomomentum (3.23) is of evident fundamental importance because it permits the
explicit construction of the Hamiltonian, symmetries, applications, etc.

The integral calculus also admits a simple isotopy with basic definitions
$\int =\int\tim\h T$ for which $\h{\int}\h{\parr}\h x=\h x$. For additional
details, one may consult [6l].


\subsection{Isofunctional analysis}

It has been proved [6j] that the
elaboration of data in RHM via ordinary and special functions and transforms
is inconsistent because not invariant under the time evolution of the
theory. This has required the isotopic lifting of functional analysis we
cannot possibly review here [6j].

We merely mention for future use that in the transition from the
two-dimensional {\sl iso-Euclidean space} with basic unit $\h I=\diag\,
({n_1}^2,{n_2}^2)$ to the {\sl iso-Gauss plane} for the characterization of
isotrigonometric functions, the isounit assumes the value $\h I_{\theta}=
n_1\tim n_2$ and  $\h I_{\phi}=n_3$ while angles assume the isotopic value
$\h{\theta}={\theta}/n_1\tim n_2$, $\hph=\phi /n_3$. This permits the
construction of the  {\sl isotrigonometric functions}

\def\t{\theta}

$$
\mb{isocos}\:\h{\theta}=n_1\tim\cos(\t /n_1\tim n_2)\:,\quad
\mb{isosin}\:\h{\t}=
n_2\tim\sin(\t/n_1\tim n_2)\:,\e{3.24}
$$

\no with corresponding {\sl isospherical coordinates}

$$
x=r\:\mb{isosin}\:\phi\:\mb{isocos}\:\t\:,
\quad y=r\:\mb{isosin}\:\phi\:\mb{isocos}\:\h{\t}\:,
\quad z=r\:\mb{isocos}\:\phi\:.\e{3.25}
$$

\no The {\sl isohyperbolic  functions} and other structures are then
constructed accordingly. Particular important for application is the
{\sl iso-Dirac delta function} $\h{\delta}(\h{x})$ which, in general, has no
longer a singularity at $\h x$, thus having intriguing conceptual and
technical implications in the possible removal of singularities
{\sl ab initio} (see [6j], for brevity).


\subsection{Lie-Santilli isotheory}

The fundamental algebraic
structure of RQM, Lie's theory, is linear, local-differential and
canonical-unitary. As such, it
is insufficient to characterize the desired nonlinear, nonlocal-integral and
noncanonical-nonunitary component of the nuclear force due to mutual
penetration of the hyperdense charge distributions of nucleons.

The primary isotopies of the original proposal [6a,6b,6d] to build HM were
those of Lie's theory, i.e., the isotopies of universal enveloping associative
algebras, Lie algebras, Lie groups, representation theory, etc. which are
today called {\it Lie-Santilli isotheory} [7].

Again, by conception and construction, the  Lie-Santilli isotheory
{\sl is not}
a new theory, but merely a {\sl new realization} of the abstract axioms of
Lie's theory. Also, recall that all Lie algebras (over a field of
characteristic zero) are known from Cartan's classification. Therefore, the
isotopies of Lie's theory cannot possibly produce new algebras, and have been
constructed instead to produce {\sl novel realizations} of known Lie algebras.

\def\po{Poincar\'e}

The main lines of the Lie-Santilli isotheory can be summarized as follows.
Let $\xi(L)$ be the universal enveloping associative algebra of an
$n$-dimensional Lie algebra $L$ with generators $X=\{X_k\}=\{{X_k}^{\dag}\}$,
unit $I$, associative product $X_i\tim X_j$, and infinite-dimensional basis
$I$, $X_k$, $X_i\tim X_j$, $i\leq j$, $X_i\tim X_j\tim X_k$, $i\leq j\leq k$,
$\ldots$ (Poincar\'e-Birkhoff-Witt theorem), and related exponentiation
$e^{iXw}=I+(i\tim X\tim w)/1!+(i\tim X\tim w)\tim(i\tim X\tim w)/2!+\ldots$,
$w\in{\cal R}$.

\def\hx{\h X}

The {\sl universal enveloping isoassociative algebra} $\h{\xi}(L)$, first
proposed in [6a,6d], is the isotopic image of  $\xi(L)$ with isounit $\h I$,
the same generators $\h X_k=X_k$ only computed in isospace, isoassociative
product $\h X_i\hh\h X_j$, infinite dimensional isobasis $\hi$, $\hx_k$,
$\hx_i\hh\hx_j$, $i\leq j$, $\hx_i\hh\hx_j\hh_s\hx_k$, $i\leq j\leq k$,
$\ldots$ (isotopic \po -Birkhoff-Witt theorem [6a,6d,7c]), and {\sl
isoexponentiation}

$$
\h e^{i\tim X\tim w}\equiv\h e^{\h i\hh\hx\hh\h w}=
\{e^{iX\tim\h T\tim w}\}\tim
\hi\:,\e{3.26}
$$

\no where $w=\{w_k\}\in{\cal R}$, and $\h w=x\tim\hi$ are the {\sl
isoparameters}. The script $\h{\xi}(L)$ rather than $\h{\xi}(\h L)$ is used in
the literature [6,7] because, when $\h I$ is no longer positive-definite, in
general $\h L \approx\hspace{-4mm}/\hspace{2mm} [\h{\xi}(L)]^-$.
This perm, its the study of a rather
intriguing {\sl unification of all simple, compact and noncompact Lie algebra
of the same dimension into one unique isoalgebra} [6j]. Note the uniqueness of
isoexponentiation (3.26) as compared to the {\sl lack} of uniqueness of the
exponentiation for $q$- and other deformations [5].

Let $L$ be the Lie algebra homomorphic to the antisymmetric algebra
$[\xi(L)]^-$ of $\xi(L)$ over a field $F(a,+,\tim)$ of real, complex or
quaternionic numbers a with familiar Lie's second theorem $[X_i,X_j]=
X_i\tim X_j-X_j\tim X_i={C_{ij}}^k\tim X_k$. The {\sl Lie-Santilli isoalgebra}
is the isospace $\h L$ with elements $\hx_k=X_k={X_k}^{\dag}$ on $\h{\cal H}$
over $\h F$ with the  {\sl isocommutation rules} [6a,6b,6d]

$$
[\hx_i\h ,\hx_j]=\hx_i\hh\hx_j -\hx_j\hh\hx_i=
{\h C_{ij}}^{\:\:\:k}\hh\hx_k\:,\e{3.27}
$$

\def\ha{\h A}
\def\hb{\h B}
\def\hc{\h C}
\def\k{\h ,}

\no whose brackets satisfy {\sl Lie's axioms in the isotopic form}
$[\ha\h ,\hb]=-[\hb\k\ha]$, $[\ha\k[\hb\k\hc]]+[\hb\k[\hc\k\ha]]+[\hc\k[\ha\k
\hb]]=0$, and the {\sl isodifferential rules} $[\ha\hh\hb\k\hc]=\ha\hh
[\hb\k\hc]+[\ha\k\hc]\hh\hb$.

Let $G$ be the (connected) Lie transformation group  characterized by the
``exponentiation" of $L$ into the elements $U(w)=e^{i\tim X\tim w}$ with
familiar laws $U(w)\tim U(w')=U(w+w')$, $U(w)\tim U(-w)=U(0)=I$. Then the
(connected) {\sl Lie-Santilli isotransformation group} $\h G$ is the
``isoexponentiation" of $\h L$ according to Eq.s (3.26) with isotopic laws

$$
\h x'=\h U(\h w)\hh\h x=\h e^{\h i\hh\hx\hh\h w}\hh\h x=
$$

$$
=\{e^{i\tim X\tim\h
T\tim w}\}\tim\hi\tim\h T\tim\h x=\{e^{i\tim X\hh\h T\tim w}\}\tim\h x\:,
\e{3.28a}
$$

$$
\h U(\h w)\hh\h U(\h w')=\h U(\h w +\h w')\:,\quad \h U(\h w)\hh\h U(-\h w)=
\h U(\h 0)=\hi\:.\e{3.28b}
$$

The nontriviality of the above isotopic theory over the conventional
formulation is then established by {\sl the appearance of the isotopic
element}
$\h T$ {\sl with an unrestricted functional dependence in the exponent of the
group structure}. This guarantees that the  Lie-Santilli isotheory has the
most general possible nonlinear, nonlocal-integral and
nonhamiltonian-nonunitary structure, although reformulated in an identical
isolinear, isolocal and isounitary form.

A main difference between the Lie theory and the covering Lie-Santilli
isotheory is that the former admits only one formulation, while the latter
admits {\sl two} formulations, one in isospace over isofields, and the other
given by its {\sl projection} in the original space.

As a general rule, {\sl the Lie and Lie-Santilli theories coincide when
formulated
in their respective spaces}, and this applies also for weights and the
representation theory. However, the projection of the latter in the space of
the  former shows deviations called {\sl mutations} which will
be illustrated shortly.

We are now equipped to indicate the preservation of the  Fermi-Dirac character
of nucleons under the simplest possible isotopy are considered that
characterized by nonunitary transforms (3.1) with a diagonal isounit $\hi$.
The problem belongs to the study of the axiom-preserving isotopies $S\h U(2)$
of $SU(2)$-spin initiated by Santilli [6h] (which are differet than the
axiom-violating deformations $SU_{p,q}(2)$ initiated in [4e]). The same
isotopies are  reformulated below apparently for the first time via general
rule (3.1), resulting in a new class of isorepresentations of $S\h U(2)$ of
rather simple construction and  effective applications.

\def\s{\sigma}

Recall that the regular (two-dimensional) representation of $SU(2)$ is
characterized by the conventional {\sl Pauli matrices} $\sigma_k$ with
familiar
commutation rules $[\s_i,\s_j]=2\tim i\tim\epsilon_{ijk}\tim\s_k$ and
eigenvalues $\s^2\tim |\psi \r =\s_k\tim\s_k\tim |\psi \r =3\tim |\psi \r $,
$\s_3\tim |\psi \r =\pm 1\tim |\psi\r $ on $\cal H$ over $\cc $.

RHM requires the construction of {\sl nonunitary images of Pauli's matrices},
which are here submitted for the first time (within the context of RHM) via
the rules

$$
\h{\s}_k=U\tim\s_k\tim U^{\dag}\:,\quad U\tim U^{\dag}=\hi\neq I\:,\e{3.29a}
$$
$$
U=\left(
\begin{array}{cc}
i\times m_{1} & 0 \\
0 & i\times m_{2}
\end{array}
\right) \; ,\quad
U^{\dagger}=\left(
\begin{array}{cc}
-i\times m_{1} & 0 \\
0 & -i\times m_{2}
\end{array}
\right)\:,
$$
$$
\hat{I}=\left(
\begin{array}{cc}
{m_{1}}^{2} & 0 \\
0 & {m_{2}}^{2}
\end{array}
\right)\:,\quad
\hat{T}=\left(
\begin{array}{cc}
{m_{1}}^{-2} & 0\\
0 & {m_{2}}^{-2}
\end{array}
\right)\:, \eqno{(3.29b)}
$$
where the $m$'s are well behaved nowhere null functions, resulting in the
{\it regular iso-Pauli matrices}
$$
\hat{\sigma}_{1}=\left(
\begin{array}{cc}
0 & {m_{1}}^{2} \\
{m_{2}}^{2} & 0
\end{array}
\right)\:,\quad
\hat{\sigma}_{2}=\left(
\begin{array}{cc}
0 & -i\times {m_{1}}^{2} \\
i\times {m_{2}}^{2} & 0
\end{array}
\right)\:,\quad
\hat{\sigma}_{3}=\left(
\begin{array}{cc}
{m_{1}}^{2} & 0 \\
0 & {m_{2}}^{2}
\end{array}
\right)\:.\eqno{(3.30)}
$$
Another realization is given by {\it nondiagonal unitary transforms}
$$
U=\left(
\begin{array}{cc}
0 & m_{1} \\
m_{2} & 0
\end{array}
\right)\:,\quad
U^{\dagger}=\left(
\begin{array}{cc}
0 & m_{2} \\
m_{1} & 0
\end{array}
\right)\:,
$$
$$
\hat{I}=\left(
\begin{array}{cc}
{m_{1}}^{2} & 0 \\
0 & {m_{2}}^{2}
\end{array}
\right)\:,\quad
\hat{T}=\left(
\begin{array}{cc}
{m_{1}}^{-2} & 0 \\
0 & {m_{2}}^{-2}
\end{array}
\right)\:,\eqno{(3.31)}
$$
\noindent with corresponding {\it regular iso-Pauli matrices}
$$
\hat{\sigma}_{1}=\left(
\begin{array}{cc}
0 & m_{1}\times m_{2} \\
m_{1}\times m_{2} & 0
\end{array}
\right)\:,\quad
\hat{\sigma}_{2}=\left(
\begin{array}{cc}
0 & -i\times m_{1}\times m_{2} \\
i\times m_{1}\times m_{2} & 0
\end{array}
\right)\:,
$$
$$
\hat{\sigma}_{3}=\left(
\begin{array}{cc}
{m_{1}}^{2} & 0 \\
0 & {m_{2}}^{2}
\end{array}
\right)\:.\eqno{(3.32)}
$$
\noindent or by more general realizations of transforms (3.29a), e.g. with
Hermitean nondiagonal isounits $\hat{I}\:.$

All iso-Pauli matrices of the above regular class verify the following
isocommutators rules and isoeigenvalue equations on $\h{\cal H}$ over
$\h{\cc}$

\def\hs{\hat{\sigma}}

$$
[\h{\s}_i\k\h{\s}_j]=\h{\s}_i\tim\h T\tim\hs_j -\hs_j\tim\h T\tim \hs_i=
2\tim i\tim\epsilon_{ijk}\tim\hs_k\:\e{3.33a}
$$

$$
\hs^{\h 2}\hh |\hps \r =
(\hs_1\hh\hs_1+\hs_2\hh\hs_2+\hs_3\hh\hs_3)\hh |\hps \r
=3\tim |\hps \r \:,\e{3.33b}
$$

$$
\hs_3\hh |\hps \r =\pm 1\tim |\hps \r \:.\e{3.33c}
$$

\no and this establishes the preservation of the  Fermi-Dirac statistics and
Pauli's exlusion principle for the nuclear realization of RHM under
consideration in this section.

We should note that realization (3.31) is the same as that constructed via the
so-called {\it Klimyk's rule} [6k], according to which

$$
\hs_k=\s_k\tim \hi\:,\quad [\hs_i\k\hs_j]=[\s_i ,\s_j]\tim\hi=2\tim i\tim
\epsilon_{ijk}\tim\s_k\tim\hi\:,\e{3.34a}
$$

$$
\hs^{\h 2}\hh |\hps \r =
3\tim |\hps \r \:, \quad \hs_3\hh |\hps \r =\pm |\hps \r \:,\e{3.34b}
$$

\no although  realization (3.29) introduced in this paper is evidently
broader.

It should be indicated for completeness that the preservation of conventional
values of spin has been specifically {\sl selected} here, because in general
the isotopies do not preserve the original eigenvalues. As an illustration,
the isoselfscalar invariance of the Hilbert space, Eq.s (3.7), implies the
existence of the following  {\it irregular iso-Pauli matrices} [6h]

$$
\hs_k=\Delta\tim\s_k\tim\hi\:,\quad [\hs_i,\hs_j]=\Delta\epsilon_{ijk}\tim
\hs_k\:,\e{3.35a}
$$

$$
\hs^{\h 2}\hh |\hps \r =
3\tim \Delta^2\tim |\hps \r \:, \quad \hs_3\hh |\hps \r =
\pm \Delta\tim |\hps \r \:,
\e{3.35b}
$$

\no where $\Delta$ is a well behaved but arbitrary non-null scalar function
usually assumed to be  $\Delta =\mb{det}\:\hi$, with evident departure from
conventional spin values.

In essence, the Fermi-Dirac character of the nucleons when members of a
nuclear structure is experimentally established and any generalization of RQM
for nuclear physics must recover this fundamental characteristics, as done
with isorepresentation (3.29).

However, the preservation of such Fermi-Dirac character is far from being
established on both theoretical and experimental grounds for the same nucleons
in more general physical conditions, e.g., in the core of a collapsing star.
The more general irregular isorepresentations of $S\h U(2)$ with  generalized
spin values have been constructed to initiate quantitative studies of the
latter more general physical conditions.


\subsection{Iso-Poincar\'e symmetry}

As it  is well known, that the Lorentz
symmetry $L(3.1)$, the Poincar\'e symmetry $P(3.1)=L(3.1)\tim T(3.1)$ and its
spinorial covering ${\cal P}(3.1)=SL(2.\cc)\tim T(3.1)$ are not
exact for isoseparation (3.19). Their isotopic images were constructed for the
first time by Santilli and called {\it iso-Lorentz symmetry} $\h L(3.1)$ [6e],
{\it iso-Poincar\'e symmetry} $\h P(3.1)=\h L(3.1)\hh\h T(3.1)$ [6g], and
{\it isospinorial covering} $\h{\cal P}(3.1)=
S\h L(2.\h{\cc})\hh\h T(3.1)$ [6i].
It was also proved that the latter isosymmetries provide indeed the
universal invariance of isoseparation (3.19). Moreover, it has been proved in
the literature that the above isosymmetries represent indeed extended,
nonspherical and deformable charge distributions under conventional values of
spin, and characterize indeed contact, nonlinear, nonlocal, nonhamiltonian and
nonunitary interactions as expected in the nuclear force.

The main characteristics of the space-time isosymmetries can be summarized as
follows. The basic isotopic structures are the field of isoreal numbers
$\h{\cal R}(\h n,+,\hh)$ and the iso-Minkowski space $\h M(\h x,\h{\eta},
\h{\cal R})$ equipped with the Tsagas-Sourlas isotopology [7f]. The
iso-Poincar\'e symmetry $\h P(3.1)$ on $\h M$ over $\h{\cal R}$ is then
constructed via the rules of the Lie-Santilli isotheory [6a,6j,7c, 7h]. This
essentially consists in preserving the {\sl conventional} generators and
parameters

$$
X=\{X_k\}=\{M_{\mu\nu}\tim p_{\alpha}\}\:,\quad M_{\mu\nu}=
x_{\mu}\tim p_{\nu}-
x_{\nu}\tim p_{\mu}\:,\e{3.36a}
$$

$$
w=\{w_k\}=\{(\t,v),a\}\in R\:,\e{3.36b}
$$

\no and by submitting to isotopies the {\sl operations} constructed of them.

\vspace{1cm}

FIGURE 4. {\bf Iso-Keplerian systems.}
As it well known, the generators $X=\{X_k\}=X^{\dag}$ of a QM space-time
symmetry represent {\sl total conserved physical quantities}, such as total
energy, total linear momentum, etc. The preservation under isotopies of the
same generators $X$ assures {\sl ab  initio} the preservation of the same
total conservation laws. Since space-time isosymmetries imply additional
interactions of contact/zero range type, we can therefore see from the outset
that  {\sl space-time isosymmetries characterize a new class of bound states,
called isokeplerian systems, for which the isotopies of Lie's theory were
proposed in the first place [6a,6b]. The new bound systems are characterized
by conventional, conserved, total physical quantities, yet with constituents
in mutual physical contact, exactly as desired for the nuclear structure.}
Computer visualization of the iso-Poincar\'e symmetry $\h P(3.1)$ then yields
the elimination of the heaviest constituent at the center, the Keplerian
nucleus, and its replacement with an arbitrary constituent, exactly as
occurring in the nuclear structure. The iso-Poincar\'e symmetry $\h P(3.1)$
studied in this section and its isospinorial covering $\h{\cal P}(3.1)$
studied in the next section are therefore expected to permit basic novel
advances in  nuclear physics studied in Sect.s 4 and 5.

\vspace{1cm}

The isotopies considered in this paper preserve conventional connectivity
properties. Therefore, connected component of the iso-Poincar\'e symmetry is
$\h P_0(3.1)=S\h O(3.1)\hh\h T(3.1)$, where $S\h O(3.1)$ is the  {\sl
connected iso-Lorentz group} [6e] and $\h T(3.1)$ is the group of
{\it isotranslations} [6g], with isotransforms on $\h M(\hx,\h{\eta},\h R)$,

$$
\h x'=\ha(\h w)\hh\h x=\ha(\h w)\tim\h T(x,\dot x,\psi ,\parr\psi ,\ldots)\tim
\h x=\tilde A(w)\tim\h x\:,
$$

$$
\ha=\tilde A\tim\hi\:,\e{3.37}
$$

\no where the first form is the mathematically correct one, the last form
being used for computation simplicity. Note that the use of conventional
linear transforms $\h x'=A(w)\tim\h x$ would now violate linearity in
isospace, besides not yielding the desired symmetry of isoseparation (3.19).

The (connected component of the) {\sl iso-Poincar\'e group} can be written in
terms of isoexponentiations (3.26) as (or can be defined by) [6g]

$$
\h P_0(3.1):\ha(\h w)=\Pi_k\h e^{iX\tim w}=(\Pi_ke^{i X\tim\h T\tim w})\tim\hi
=\tilde A(w)\tim\hi\:.\e{3.38}
$$

The preservation of the original dimension is ensured by the {\sl isotopic
Baker-Campbell-Hausdorff Theorem} [6a]. It is easy to see that structure
(3.38) forms a connected {\it Lie-Santilli transformation isogroup}.

To identify the isoalgebra $\h p_0(3.1)$ of $\h P(3.1)$ we use the
isodifferential calculus (Sect. 3.7) and isolinear momentum (3.23) which
yield the {\it isocommutation rules} [6g]

\def\hm{\hat M}
\def\a{\alpha}
\def\b{\beta}

$$
[\hm_{\mu\nu}\k\hm_{\alpha\beta}]=i(\h{\eta}_{\nu\a}\hm_{\mu\b}-
\h{\eta}_{\mu\a}\hm_{\nu\b}-\h{\eta}_{\nu\b}\hm_{\mu\a}+\h{\eta}_{\mu\b}
\hm_{\a\nu})\:,\eqno{(3.39a)}
$$

$$
[\hm_{\mu\nu}\k\h p_{\a}]=i(\h{\eta}_{\mu\a}\h p_{\nu}-\h{\eta}_{\nu\a}
\h p_{\mu})\:,\quad[\h p_{\a}\k\h p_{\b}]=0\:,\e{3.39b}
$$

\no where $[A\k B]=A\tim\h T(x,\psi,\ldots)\tim B-B\tim\h T(x,\psi ,\ldots)
\tim A$.

The {\it iso-Casimir invariants} are then lifted into the forms [loc.cit.]

$$
C^{(0)}=\hi(x,\dot x, \psi ,\parr\psi ,\ldots)=\h T^{-1}\:,\e{3.40a}
$$

$$
C^{(1)}=\h p^{\h 2}=\h p_{\mu}\hh\h p^{\mu}=\h{\eta}^{\mu\nu}
\h p_{\mu}\hh \h p_{\nu}\:,\e{3.40b}
$$

$$
C^{(3)}=\h W_{\mu}\hh\h W^{\mu}\:,\quad \h W_{\mu}=\epsilon_{\mu\a\b\rho}
\hm^{\a\b}\hh\h p^{\rho}\:.\e{3.40c}
$$

The local isomorphism $\h p_0(3.1)\approx p_0(3.1)$ is ensured by the
positive-definiteness of $\h T$. Alternatively, the use of the generators
in the form ${\hm^{\mu}}_{\:\:\nu}=x^{\mu}\tim p_{\nu}-x^{\nu}\tim p_{\mu}$
yields the
{\sl conventional} structure constants under a {\sl generalized} Lie product,
as one can verify via the use of properties (3.21). The above local
isomorphism is sufficient, per s\'e, to guarantee the axiomatic consistency
of RHM.

The main components of $\h P(3.1)$ are the following:


\vspace{0.5\baselineskip}

{\bf 3.10.A. Isorotations,} which are the space components
$S\h O(3)$[6e,6g,6h]. They can be computed from isoexponentiations (3.38) and
the space components $\h T_{kk}$ of the isotopic element in diagonal form,
$\h T=\diag\,(T_{\mu\mu})$, $T_{\mu\mu}={\h T_{\mu}}^{\:\:\nu}$,
yielding the {\sl isorotations in the (x,y)-plane}

$$
x'=x\tim\cos({\h T_{11}}^{\:\:\: {1\over 2}}\tim{\h T_{22}}^{\:\:\:{1\over 2}}
\tim\t_3)-
\h y\tim{\h T_{11}}^{\:\:\:-{1\over 2}}\tim{\h T_{22}}^{\:\:\:{1\over 2}}
\tim\sin(
{\h T_{11}}^{\:\:\:{1\over 2}}\tim{\h T_{22}}^{\:\:\:{1\over 2}}\tim \t_3)\:,
\e{3.41a}
$$

$$
y'=\h x\tim{\h T_{11}}^{\:\:\:{1\over 2}}\tim{\h T_{22}}^{\:\:\:-{1\over 2}}
\tim\sin({\h T_{11}}^{\:\:\:{1\over 2}}\tim{\h T_{22}}^{\:\:\:{1\over 2}}
\tim\theta_3)+
\h y\cos({\h T_{11}}^{\:\:\:{1\over 2}}\tim{\h T_{22}}^{\:\:\:{1\over 2}}
\tim\theta_3)\:,
\eqno{(3.41b)}
$$

\no  (see [6k] for general isorotations in all three Euler angles).

As one can easily verify, isorotations (3.41) leave invariant all infinitely
possible ellipsoidical deformations of the sphere

$$
r^{\h 2}=x{\h T_{11}}x+y{\h T_{22}}y+z{\h T_{33}}z=R\:,\eqno{(3.42)}
$$

\no thus  confirming the achievement of a representation of the deformation
theory via a covering of Lie's theory, as needed for a quantitative
representation of the historical hypothesis of Fig. 1.

\vspace{1cm}

FIGURE 5. {\bf Isosphere.} A central objective of RHM (from which the new
mechanics derived its
name [6b]) is the representation of hadrons as they are expected to be in the
physical reality: {\sl extended, nonspherical and deformable charge
distribution}. In nuclear physics, the representation of these characteristics
must be achieved under the condition of preserving conventional values of
spin.
The achievement of this dual objective is geometrically established by the
notion depicted in this figure, the {\it isosphere}, which maps all infinitely
possible ellipsoidical shapes into the {\sl perfect sphere} $r^{\h 2}=(r^t\tim
\delta\hh r)\tim\hi$ in the {\it iso-Euclidean spaces} $\h E(\h r,\h{\delta},
\h{\cal R})$, $\h r=\{\h r^k\}=\{r^k\}$, $\h{\delta}=\h T_{s}\tim\delta$,
$\delta=\diag\,(1,1,1)$, $\h T_s=\diag\,(\h T_{11},\h T_{22},\h T_{33})$,
${\hi_s}={\h T_s}^{\:\:-1}$ [6j]. In turn, the reconstruction of the perfect
spheridicity assures the preservation of the exact rotational symmetry,
$\h O(3)\approx O(3)$ and $S\h U(2)\approx SU(2)$, and , consequently, of
conventional values of the
orbital and intrinsic angular momenta. In fact, the
lifting of the semiaxes of the perfect sphere into those of spheroidal
ellipsoids, $1_k\arr\h T_{kk}$, when the related units are lifted of the
{\it inverse} amounts, $1_k\arr{\h T_{kk}}^{\:\:\:-1}$, implies the
preservation of
the perfect sphericity. The novel model of nuclear structure permitted by the
iso-Poincar\'e symmetry (Fig. 4) is therefore based on {\sl nucleons
represented
as isospheres}, which are perfect sphere when represented in isospace $\h E$,
but when projected in our space $E$ are given by all infinitely possible
spheroidal ellipsoids, exactly as desired for the historical hypothesis of
Fig. 1.

\vspace{1cm}

{\bf 3.10.B Iso-Lorentz boosts,} which can be written explicitly in the
(3,4)-plane [6e]

$$
{x^1}'=x^1\:,\quad {x^2}'=x^2\:,\eqno{(3.43a)}
$$

$$
{x^3}'=x^3\tim\sin h({\h T_{33}}^{\:\:\:{1\over 2}}
\tim{\h T_{44}}^{\:\:\:{1\over 2}}\tim v)
-x^4\tim{\h T_{33}}^{\:\:\:-{1\over 2}}\tim{\h T_{44}}^{\:\:\:{1\over 2}}
\tim\cos h
({\h T_{33}}^{\:\:\:{1\over 2}}\tim{\h T_{44}}^{\:\:\:{1\over 2}}\tim v)=
$$

$$
=\h{\gamma}\tim(x^3-{\h T_{33}}^{\:\:\:-{1\over 2}}
\tim{\h T_{44}}^{\:\:\:{1\over 2}}\tim\h{\beta}
\tim x^4)\:,\eqno{(3.43b)}
$$

$$
{x^4}'=-x^3\tim{\h T_{33}}^{\:\:\:{1\over 2}}2
\tim{c_0}^{-1}{\h T_{44}}^{\:\:\:-{1\over 2}}\tim
\sin h({\h T_{33}}^{\:\:\:{1\over 2}}\tim{\h T_{44}}^{\:\:\:{1\over 2}}
\tim v)
+x^4\tim\cos h({\h T_{33}}^{\:\:\:{1\over 2}}
\tim{\h T_{44}}^{\:\:\:{1\over 2}}\tim v)=
$$

$$
=\h{\gamma}\tim(x^4-{\h T_{33}}^{\:\:\:-{1\over 2}}
\tim{\h T_{44}}^{\:\:\:-{1\over 2}}
\tim\h{\beta}\tim x^3)\:,\eqno{(3.43c)}
$$

\no where

$$
\h{\beta}=(v_k\tim\h T_{kk}\tim v_k/c_0\tim\h T_{44}\tim c_0)^{1\over 2}\:,
\eqno{(3.44a)}
$$

$$
\h{\gamma}=(1-\h{\beta}^{\h 2})^{-1/2}\:.\eqno{(3.44b)}
$$

Note that the above isotransforms are {\sl nonlinear} (in $x,\dot x,\psi,\parr
\psi ,\ldots$), {\sl nonlo\-cal-integral} (e.g., because the factor
$\h{\Gamma}$
in (3.3) can be of the type $\exp\int dv\hps^{\dag}\hps$ representing
precisely
the overlapping of the wavepackets of the constituents) {\sl and nonunitary}
(because the isoexponentiations  (3.38) are indeed nonunitary in $\cal H$),
precisely as desired, yet they are formally similar to the Lorentz transforms,
as expected from their isotopic character. This also confirms the local
isomorphism $S\h O(3.1)\approx SO(3.1)$ [6e].

\vspace{0.5\baselineskip}


{\bf 3.10.C. Isotranslations,} which can be written [6g]

$$
x'=(\h e^{ip\tim a})\hh x =x+a\tim A(x,\ldots)\:,\quad p'=(\h e^{ip\tim a})\hh
p=p\:,\eqno{(3.45a)}
$$

$$
A_{\mu}={\h T_{\mu\mu}}^{\:\:\:1/2}+a^{\alpha}[{\h T_{\mu\mu}}^{\:\:\:1/2}
\k p_{\alpha}]
/1!+\ldots\eqno{(3.45b)}
$$

\vspace{0.5\baselineskip}
{\bf 3.10.D. Isoinversions,} expressible in the forms

$$
\h{\pi}\hh x=\pi\tim x=(-r,x^4)\:,\quad\h{\tau}\hh x=\tau\tim x=(r,-x^4)\:,
\eqno(3.46)
$$

\no where $\h{\pi}=\pi\tim\hi$, $\h{\tau}=\tau\tim\hi$, and $\pi$, $\tau$ are
the conventional inversion operators; and the

\vspace{0.5\baselineskip}

{\bf 3.10.E. Isoscalar transforms,} which are the new transforms

$$
\hi\arr\hi '=n^2\tim\hi\:,\quad \h{\eta}\arr\h{\eta}'=n^{-2}\tim\h{\eta}\:,
\eqno{(3.47)}
$$

\no leaving invariant the conventional or isotopic separation, Eq.s (3.19).

Thus, {\sl the iso-Poincar\'e symmetry is 11-dimensional}, i.e., it has the
10 conventional parameters, plus the parameters $n^2$ of isotransforms (3.47).
Needless to say, the latter new invariance can also be defined for the
conventional Poincar\'e symmetry which, as such, also acquires 11 dimensions.

The isospinorial covering $\h{\cal P}(3.1)$ will be identified in the next
section. The construction of the {\sl iso-Galilean symmetry} $\h G(3.1)$ via
the isotopies of conventional techniques on constractions is an instructive
exercise for the interested reader (see also [6k] for an explicit
realization).


\subsection{Isospecial relativity}

On rigorous scientific grounds, the
validity of the conventional formulation of the special relativity is nowadays
restricted to motion of particles or electromagnetic waves {\sl in vacuum}
with constant maximal causal speed $c_0$. This is due to the fact that, on one
side, it is known since the past century (see, e.g., the studies by Lorentz
[12a] and their review by Pauli [12b]) that electromagnetic waves propagate
within physical media with a locally varying speed
$c=c_0/n_4(x,\ldots)\l c_0$,
as it is the case in our atmosphere, water, plastic, glasses, oil, etc.

On the other side, photons propagating within certain guides with speeds
$c=c_0/n_4(x,\ldots)\r c_0$ have been experimentally measured [13a,13b], and
large masses have been measured in astrophysics to be expelled at speeds
bigger than $c_0$ [13c,13d,13e].

Moreover, wave solutions of {\sl conventional} relativistic equations with
{\sl arbitrary} speeds have been recently detected in [13f]. Thus, nowadays
the speed of electromagnetic waves is no longer a ``universal constant" but a
local quantity smaller or bigger than $c_0$ which assumes the constant value
$c_0$ only in vacuum.

It is evident that the special relativity in its current formulationis
{\sl inapplicable} (and not ``violated") for locally varying speeds
$c=c_0/n_4(x,\ldots)$. In addition to the evident loss of the Lorentz and
Poincar\'e symmetries, the insistence of the applicability of the special
relativity under conditions for which it was not meant for, e.g., in water,
leads to inconsistencies, such as: the assumption of the speed of light
$c=c_0/n_4$ in water as the maximal causal speed implies the violation of the
principle of causality because electrons can travel in water faster than the
speed of light (Cerenkov light); the assumption of the speed of light $c_0$
{\sl in vacuum} as the maximal causal speed {\sl in water} to salvage the
principle of causality implies the violation of the relativistic addition of
speeds for which the sum of two speeds of light in water does not yield the
speed of light; and other inconsistencies [6k].

Moreover, the special relativity is also known not to be applicable for the
description of deformations, as needed for the historical hypothesis
of Fig. 1,
and can  characterize only linear, local-differential and Hamiltonian-unitary
systems, while a primary objective of these studies is a quantitative
treatment of the nonlinear, nonlocal and nonunitary component expected in the
nuclear force.

The {\it isospecial relativity} was proposed by Santilli [6e,6g,6k] for: the
form-invariant description of arbitrary speeds $c=c_0/n_4$;
the characterization of extended-deformable shapes of
particles; and the form-invariant description of nonlinear, nonlocal and
nonunitary interactions. The isospecial relativity is characterized by the
axioms of the conventional formulation merely realized in isominkowski
space $\h M$ over $\h{\cal R}$ under the iso-Poincar\'e invariance
$\h P(3.1)$. As such, the special and isospecial relativity coincide at the
abstract level by conception and construction, as it is the case for all
other aspects of RHM.

\vspace{1cm}

FIGURE 6. {\bf a) Light cone in physical space, b) light ``cone'' in physical
media, c) isolight cone in isospace.} The conventional light cone is well
defined only in empty space where light has the constant speed $c_0$ (Fig. a).
Within physical media the speed of electromagnetic waves is however a local
variable, thus implying evident deformations of the  conventional light cone
(Fig. B). The iso-Lorentz symmetry (6e) maps the latter deformed surface into
the {\it isolight cone}, which is the perfect cone in iso-Minkowskian space
$\h M(\h x,\h{\eta},\h R)$ (Fig. c). In a way similar to the  isosphere, we
have the deformation of the light cone axes $!-{\mu}\arr\h T_{\mu\mu}$ while
the corresponding units are deformed of the  {\sl inverse} amount $1_{\mu}
\arr{\h T}_{\mu\mu}^{\:\:\:-1}$, thus preserving the original characteristics
of a perfect cone. Such a preservation is then the geometric foundation of
the local isomorphism $S\h O(3.1)\approx SO(3.1)$. The axiom-preserving
character of the isotopy of the light cone is so strong that even the
characteristic angle of the cone remains the conventional one, i.e.,
{\sl the maximal causal speed in isospace $\h M(\h x,\h{\eta},\h R)$ remains
the speed of light $c_0$ in vacuum} [6g] (it should be noted that the proof of
this property requires, for consistency, the use of the isotrigonometric and
isohyperbolic functions we cannot review here for brevity [6j]). This
establishes the capability of all problems addressed in this paper to be
formulated in a way compatible with the special relativity, only realized in
isospace $\h M$.

\vspace{1cm}

Variable speeds of electromagnetic waves propagating within inhomogeneous and
anisotropic physical media are geometrically represented in a direct way via
the isoseparation on  $\h M$ for
$\h T=\diag\,(\{{n_1}^{-2},{n_2}^{-2},{n_3}^{-2}\},{n_4}^{-2})$,
$n_{\mu}\neq 0$,

$$
x^{\h 2}=[x^{\mu}\h{\eta}_{\mu\nu}(x,\dot x,\ldots)x^{\nu}]\tim\hi=
$$

$$
=(xx/{n_1}^2+yy/{n_2}^2+zz/{n_3}^2-tt\tim{c_0}^2/{n_4}^2)
\tim\hi\in\h{\cal R}\:.
\eqno{(3.48)}
$$

\no Its evident universal invariance is given by the iso-Poincar\'e symmetry
$\h P(3.1)$. Propagation within homogeneous and isotropic media is expressed
by the new invariance (3.20),

$$
x^{\h 2}=[x^{\mu}\h{\eta}_{\mu\nu}(x,\dot x ,\ldots)x^{\nu}]\tim\hi =
$$

$$
=(xx/n^2+yy/n^2+zz/n^2-tt\tim{c_0}^2/n^2)\tim(n^2\tim I)\equiv x^{\h 2}\:,
\eqno{(3.49)}
$$

\no which is the fundamental symmetry underlying the waves of arbitrary
speeds of ref. [13f].

Note that, despite the formal  identity $x^{\h 2}\equiv x^2$, {\sl the use of
the iso-Poincar\'e symmetry is necessary for the invariance under arbitrary
speeds} $c=c_0/n$. In turn, this implies the activation of the entire isotopic
formalism of this section. The nontriviality of the $\h P(3.1)$-invariance is
then reflected in the appearance of the function $n^2(x,\ldots)$ in the
{\sl arguments} of the isorotations, isolorentz boosts, isotranslations,
etc., Eq.s (3.41)-(3.47).

One of the first implications of the isospecial relativity is that of
permitting the {\sl representation of locally varying speeds of light via the
conventional, abstract axioms of the special relativity} [6e,6g]. This is
achieved via the reconstruction of $c_0$ as the unique and universal maximal
causal speed in isospace, while its projection in our space-time can assume
any possible speed. In fact, jointly with the change $c_0\arr c_0/n_4$ the
unit changes by the inverse amount $1\arr n_4$, thus preserving the original
value $c_0$. In this way $c_0$ is a
``universal constant" only in isospace $\h M$, while its {\sl projection} in
conventional space-time acquires the local form $c_0/n_4$.

The compatibility of the isospecial relativity with deformable shapes
(indicated from the title of the first proposal [6e]) is evident from the
unrestricted character of the functional dependence of the isounit.
The nonlinear, nonlocal and nonunitary characters
are equally evident from the structure of the iso-Poincar\'e symmetry.

Intriguingly, we can say that {\sl the special relativity is universally
applicable only in isospace over  isofields}, because only in this latter case
we have one single unique and universal causal speeds $c_0$, all possible
deformed light cones are reduced to the perfect cone in isospace, and
nonlinear, nonlocal and nonunitary interactions are identically rewritten in
their isolinear, isolocal and isounitary form.


\subsection{Isotopic dynamical equations}

The {\sl fundamental isorelativistic
equations} are uniquely identified by the iso\-Poincar\'e symmetry via its
iso-Casimir invariants (3.40) and related isorepresentation theory which we
cannot possibly study here for brevity (see ref. [6k] for initial studies).
The basic equation is the {\sl second-order isorelativistic equation} which is
given by the isoinvariant (3.40b) implemented with the conventional minimal
coupling rule to an external electromagnetic field with four-potential
$\h A_{\mu}(x)$, and realized in terms of the isodifferential calculus (3.21)

$$
\{[\h p_{\mu}+i\tim e\tim\h A_{\mu}]\hh [\h p^{\mu}+i\tim e\tim\h A^{\mu}]
+\h m^{\h 2}\}\hh |\hps\r=
$$

$$
=\{\h{\eta}^{\mu\nu}\tim[\h p_{\mu}+ i\tim e\tim\h A_{\mu}]\tim\h T\tim
[\h p_{\mu}+i\tim e\tim\h A_{\nu}]+(m\tim m)\tim\hi]\tim\h T\tim |\hps\r=
$$

$$
=\{\h{\eta}^{\mu\nu}[-i\h{\partial}_{\mu}+i\tim e\tim A_{\mu}]\tim
[-i\h{\parr}_{\nu}+i\tim e\tim A_{\nu}]+m^2\}\tim |\hps\r=
$$

$$
=\{{\hi^{\mu}}_{\:\:\alpha}\h{\eta}^{\a\nu}[-i{\h T_{\mu}}^{\:\:\gamma}\tim
\parr_{\gamma}+i\tim e\tim A_{\mu}]\tim[-i{\h T_{\nu}}^{\:\:\delta}\tim
\parr_{\delta}+i\tim e\tim A_{\nu}]+m^2\}\tim |\hps\r=0\:.\eqno{(3.50)}
$$

A solution for the case of null external field and isounits averaged to
constant diagonal elements ${n_{\mu}}^2$ is given by the {\sl isoplane wave}
($c_0=1$)

$$
\psi(x)=e^{i{\hi^{\mu}}_{\:\:\nu}\tim p_{\mu}\tim x^{\nu}}\:,\eqno{(3.51)}
$$

\no which does reproduce isoinvariant (3.40b) for constant $p$'s and $n$'s.

For completeness, we quote here the nonrelativistic equations [6k,6l]

$$
i\h{\parr}_t\hps =i\h T_t\parr_t\hps =\h H\hh_s\hps=\h H\tim\h T_s\tim
\hps=
$$

$$
=\h E\hh_s\hps=(E\tim\hi_s)\tim\h T_s\tim\hps=E\tim\hps\:,\eqno{(3.52a)}
$$

$$
\hps(t,r)=\{\h e^{iH\tim t}\}\hh_s\psi(0,r)=\{e^{i\h H\tim\h T_s\tim t}\}\tim
\psi(0,r)\:,\eqno{(3.52b)}
$$

$$
i\h d\h A /\h d t=i\hi_td\h A/dt=\h A\hh_s\h H-\h H\hh_s\h A=\h A\tim\h T_s
\tim\h H-\h H\tim\h T_s\tim\ha\:,\eqno{(3.52c)}
$$

$$
\ha(t)=\{\h e^{i\tim\h H\tim t}\}\hh_s\ha(0)\hh_s\{\h e^{it\tim\h H}\}=
\{e^{i\h H\tim\h T_s\tim t}\}\tim\ha(0)\tim\{e^{it\tim\h T_s\tim\h H}\}\:,
\eqno{(3.52d)}
$$

$$
\h p\hh_s\hps=\h p\tim\h T_s\tim\hps=-i\h{\nabla}_k\hps=-i{\h T_k}^{\:\:i}
\nabla_i\hps\:,\eqno{(3.52d)}
$$

$$
[\h p_i\k\h r^j]=\h p_i\hh_s\h r^j-\h r^j\hh_s\h p_i=-i{\delta_i}^jI\:,
\quad [\h p_i\k\h p_j]=[\h r^i\k\h r^j]\equiv 0\:,\eqno{(3.52e)}
$$

$$
\hi_i=\hi_t(t,r\hps ,\ldots)={n_4}^2\tim\h{\Gamma}_t(t,r,\hps ,\ldots)=
{\h T_t}^{\:\:-1}\r 0\:,\eqno{(3.52f)}
$$

$$
\h T_s=\h T_s(t,r,\hps ,\ldots)=\mb{diag}({n_1}^2,{n_2}^2,{n_3}^2)\tim
\h{\Gamma}_s(t,r,\hps ,\ldots)\r 0={\hi_s}^{\:\:-1}\:.\eqno{(3.52g)}
$$

\no with isoplane-wave solution

$$
\hps(t,r)=e^{i\tim(p_k\tim{n_k}^2\tim r_k-E\tim{n_4}^2\tim t)}\:,\eqno{(3.53)}
$$

\no which coincides with (3.51) as in the  conventional case.

One should note the compatibility of the relativistic and nonrelativistic
equations, the ``decoupling" in the latter of the isounit into i8s space and
time components, as well as the isounitary structure of the time evolution in
finite form. The proof that the above dynamical equations are  indeed
form-invariant under their respective relativistic and nonrelativistic
isounitary symmetries is an instructive exercise for the interested reader.

\section{Exact representation of nuclear \hfil\break magnetic moments}

Once the new formalism of RHM is known, the {\sl exact} representation of the
total magnetic moments of few-body nuclei becomes straightforward. Its
simplicity and exact character should then be compared with the truly complex
calculations of ref. [2] via the conventional relativistic/Bethe-Salpeter
theories and its lack of exact character.

The most effective derivation is that via the {\it iso-Dirac equation}, i.e.,
the isotopies of the conventional Dirac equation [14a] originating from the
linearization of the  second-order isorelativistic equation (3.50). This
linearization has been studied by a number of authors (see [6k], Ch. 10, for
details and references), although in its general form it implies a {\it
mutation of sp6in} which, quite intriguingly, was first discovered by Dirac
himself [14b,14c], although without his knowledge of the essential isotopic
structure of his own generalized equation [6k].

In this paper we have to re-inspect the derivation of the iso-Dirac equation
and introduce a new form specifically intended to represent the mutation of
the intrinsic magnetic moments of nucleons while preserving their angular
momentum and spin, as necessary for a study of the historical hypothesis
(Sect. 1). For other generalizations representing the (constant, yet)
{\it anomalous} magnetic moments of nucleons one may inspect ref. [14d] and
papers quoted therein.

The linearization of second-order isoinvariant (3.40b) requires a {\it
composite isounit} characterized by the tensorial product of two nonunitary
transforms, one acting in the {\it orbital} component and on in the {\it spin}
part, resulting in the 8-dimensional total isounit,

$$
(U^{\orb}\otimes U^{\spin})\times(U^{\orb}\otimes U^{\spin})^{\dag}
=\hi^{\tot}=
$$

$$
=\hi^{\orb}\otimes\hi^{\spin}=
(\h T^{\orb})^{-1}\otimes(\h T^{\spin})^{-1}\r 0\:.\eqno{(4.1a)}
$$

$$
\hi^{\orb}=\diag\,({n_1}^2,{n_2}^2,{n_3}^2,{n_4}^2)\:,\quad n_{\mu}\neq 0\:,
\eqno{(4.1b)}
$$

$$
\hi^{\spin}=
\diag\,({m_1}^2,{m_2}^2,-{m_1}^2,-{m_2}^2)\:,\quad n_{\mu}\neq 0\:,
\eqno{(4.1b)}
$$

$$
U^{\spin}=\pmatrix{0&W_{2\tim 2}\cr {W^{\dag}}_{2\tim 2}&0}\:,\quad
{U^{\dag}}^{\spin}=\pmatrix{0&{W^{\dag}}_{2\tim 2}\cr W_{2\tim 2}&0}\:,
\eqno{(4.1c)}
$$

$$
W_{2\tim 2}=\pmatrix{0&m_1\cr m_2&0}\:,\quad{W^{\dag}}_{2\tim 2}=
\pmatrix{0&m_2\cr m_1&0}\:,\eqno{(4.1d)}
$$

$$
{{\hi}^{\spin}}_{\hspace{5mm}2\tim 2}= W_{2\tim 2}\tim{W^{\dag}}_{2\tim 2}=
\pmatrix{{m_1}^2&0\cr 0&{m_2}^2}\:,
$$

$$
{{\h T}^{\spin}}_{\hspace{5mm}2\tim 2}=
\pmatrix{{m_1}^{-2}&0\cr 0&{m_2}^{-2}}\:,\eqno{(4.1e)}
$$

$$
{ {\hi}^{d\:\spin} }_{\hspace{7mm}2\tim 2}=
-{ {\hi}^{\dag\spin} }_{\hspace{7mm}2\tim 2}=
-{W^{\dag}}_{2\tim 2}\tim W_{2\tim 2}=\pmatrix{-{m_1}^2&0\cr 0&-{m_2}^2}\:,
$$

$$
{ {\h T}^{d\:\spin} }_{\hspace{7mm}2\tim 2}
=\pmatrix{-{m_1}^{-2}&0\cr 0&-{m_2}^{-2}}\:.
\eqno{(4.1f)}
$$

The linearization of Eq.s (3.50) can then be written

\def\s{\sigma}

$$
({\h{\eta}}^{\rho\s}\tim\h p_{\rho}\hh\h p_{\nu}+\h m^2)\hh\hps=\eqno{(4.2)}
$$

$$
=(\h{\eta}^{\mu\nu}\tim\h{\gamma}_{\mu}\tim{\h T}^{\spin}\tim\h p_{\nu}\tim
\h T^{\orb}-i\tim m)\tim{\h T}^{\tot}
(\h{\eta}^{\a\beta}\tim\h{\gamma}_{\a}\tim
\h T^{\spin}\tim\h p_{\beta}\tim\h T^{\orb}+i\tim m)\tim\psi=
$$

$$
=[\h{\eta}^{\mu\nu}\tim\h{\eta}^{\a\beta}\tim{1\over 2}\tim
(\h{\gamma}_{\mu}\tim{\h T}^{\spin}\tim
\h{\gamma}_{\a}\tim\h T^{\spin}+
$$

$$
+\h{\gamma}_{\a}\tim\h T^{\spin}\tim
\h{\gamma}_{\mu}\tim\h T^{\spin})\tim\h p_{\nu}\tim\h T^{\orb}\tim\h p_{\beta}
\tim\h T^{\orb}+m\tim m]\tim\hps=0\:,
$$

\no resulting in the following form of the {\it iso-Dirac equation}
(apparently introduced here for the first time)

$$
(\h{\eta}^{\mu\nu}\tim\h{\gamma}_{\mu}\hh^{\spin}
\h p_{\nu}-i\tim\h m^2)\hh^{\orb}
\hps =
$$

$$
=(\h{\eta}^{\mu\nu}\tim\h{\gamma}_{\mu}\tim\h T^{\spin}\tim\h p_{\nu}\tim
\h T^{\orb}-i\tim m\tim m)\tim\hps =0\:.\eqno{(4.3)}
$$

\no The {\sl isogamma matrices} defined by

$$
\{\h{\gamma}_{\mu}\k\h{\gamma}_{\a}\}=\h{\gamma}_{\mu}\tim\h T^{\spin}\tim
\h{\gamma}_{\a}\tim\h T^{\spin}+\h{\gamma}_{\a}\tim\h T^{\spin}\tim
\h{\gamma}_{\mu}
\tim\h T^{\spin}=2\h{\eta}_{\mu\nu}\:,\eqno{(4.4)}
$$

\no and admit the explicit realization

$$
\h{\gamma}_{\mu}=({{\h T}_{\mu\mu}}^{\:\:\:\orb})^{1/2}
\tim U^{\spin}\tim{\gamma}_{\mu}
\tim{U^{\dag}}^{\spin}\tim\hi^{\spin}\:,\eqno{(4.5a)}
$$

$$
\h{\gamma}_k=({\h T_{kk}}^{\:\:\:\orb})^{1/2}\tim\gamma_k\tim\hi^{\spin}=
{{\h T}_{kk}}^{\:\:\:1/2}\tim\pmatrix{0&\h{\s}_k\cr {{\h{\s}}^d}_{\:\:k}&0}
\tim\hi^{\spin}\:,\eqno{(4.5b)}
$$

$$
\h{\gamma}_4=({\h T_{44}}^{\:\:\:\orb})^{1/2}\tim\gamma_4\tim\hi^{\spin}=
({{\h T}_{kk}}^{\:\:\:\orb})^{1/2}\tim
\pmatrix{ {I^{\spin}}_{2\tim 2}&0\cr 0 &
{ {\hi}^{d\:\spin} }_{\hspace{7mm}2\tim 2} }\tim\hi^{\spin}
\:,\eqno{(4.5c)}
$$

$$
\h{\s}_k=W\tim\s_k\tim W^{\dag}\:,\quad{{\h{\sigma}}^d}_{\:\:k}=
-{{\h{\s}}^{\dag}}_{\:\:k}=
-W^{\dag}\tim\s_k\tim W\:.\eqno{(4.5d)}
$$

\no with a simple extension to the minimal coupling rule hereon tacitly
implied.

The nontriviality of the isotopy is then established by the fact that the
isotopic elements $\h T_{\mu\mu}$ enter into the structure of the  isotopic
gamma matrices.

As one can see, the conventional Dirac equations is defined on conventional
Minkowski space with basic unit $I=\diag\,(\{1,1,1\}, 1)$, thus
characterizes
the perfect and rigid sphere $\{1,1,1\}$ mowing in vacuum, $n_4=1$. The above
isodirac equation represents instead all infinitely possible ellipsoidical
deformations of the perfect sphere with semiaxes ${n_1}^2$, ${n_2}^2$,
${n_3}^2$
under the volume preserving condition (2.1),
${n_1}^2\tim {n_2}^2\tim{n_3}^2=1$,
while propagating within a physical media with index of refraction
$n_4\neq 1$.

First, it is important to verify that, despite the alteration of the shape
of the charge
distribution, the {\sl values of the angular momenta are conventional}. This
is easily
established by the fact that the isodirac equation (4.3) characterizes the
following isotopic $S\h O(3)$ algebra (where all products are referred to the
orbital isotopic product)

$$
\h O(3)\::\quad\h M_k={1\over 2}\epsilon_{kij}\h{\h r}^i\h{\hh}\h p_j\:,
\eqno{(4.6a)}
$$

$$
[\h M_i\k\h M_j]=\h M_i\tim\h T\tim\h M_j -\h M_j\tim\h M_i=\epsilon_{ijk}\tim
\h M_k\:,\eqno{(4.6b)}
$$

$$
{\hat M}^{\hat 2}\hat{\times}\hat{\psi}=
\hat{M}_k\times\hat{T}\times\hat{M}^k\times\hat T\tim\hat{\psi}=
m(m+1)\times\hat{\psi}\:,\eqno{(4.6c)}
$$

$$
\h M_3\tim\hps =\h M_k\tim\h T\tim\hps =(\pm m)\tim\hps\:,\quad
m=0,1,2,\ldots\:.
\eqno{(4.6d)}
$$

\no This assures that Eq.s (4.3) characterize conventional eigenvalues of the
angular momentum.

Second it is easy to see that the $S\h U(2)$ spin algebra on the isofield
$\h{\cc}=\h{\cc}(\h c, +,\hh)$ as characterized by the above
isodirac equation has a {\sl generalized} structure, yet {\sl conventional}
eigenvalues as desired. In fact, we have the following expressions in terms of
spin isoproducts

$$
S\h U(2)\::\quad\h S_k={1\over 2}\epsilon_{kij}\h{\gamma}_i\hh\h{\gamma}_j\:,
\eqno{(4.7a)}
$$

$$
[\h S_i\k\h S_j]=\h S_i\tim\h T\tim\h S_j -\h S_j\tim\h S_i=\epsilon_{ijk}\tim
\h S_k\:,\eqno{(4.7b)}
$$

$$
\h S^{\h 2}\hh\hps=\h S_k\tim\h T\tim\h S^k\tim\h T\tim\hps=(3/4)\tim\hps\:,
\eqno{(4.7c)}
$$

$$
\h S_3\tim\hps =\h S_k\tim\h T\tim\hps =(\pm 1/2)\tim\hps\:,
\eqno{(4.7d)}
$$

\no which constitute a $4\tim 4$ extension of results (3.33). This assures the
characterization of conventional spin, with consequential preservation of
Pauli's exclusion principle.

The combined generators $M=(M_{\mu\nu})$, $M_{ij}=\epsilon_{ijk}S_k$, $M_{k4}=
i\h{\gamma}_k\hh\h{\gamma}_4$ then characterize the
{\it isospinorial covering}
$S\h L(2.\h{\cc})$ of the iso-Lorentz algebra $\h L(3.1)$. The study of the
isocommutation rules and local isomorphism $S\h L(2.\h{\cc})\approx SL(2.\cc)$
is left to the interested reader, jointly with the isotopies of the remaining
aspects of Dirac's theory [6k].

Iso-Dirac equation (4.3) provides the desired two generalized expressions which
are needed for a fit of the experimental data on nuclear magnetic moments.
First, Eq. (4.3) implies the following desired {\it mutation of the spinorial
transformation law}, first identified by Santilli in ref. [15]

$$
\hps '=\h R(\theta_3)\hh\hps =e^{i\gamma_1\gamma_2\h{\theta}_3/2}\hps\:,\quad
\h{\theta}=\theta /n_1n_2\:,\eqno{(4.8)}
$$

\no Then, a simple isotopy of the conventional case yields the desired {\sl
mutation of the magnetic moment of nucleons} (see [6k], Ch. 10. for details)

$$
\h{\mu}_N=\h{\mu}_N(\mu_N,{n_{\mu}}^2)=\mu_N\tim n_4/n_3\:.\eqno{(4.9)}
$$

\no first proposed in [6b]. Sect. 4.20, p. 803.

In summary, {\sl RHM can represent in first-quantization (and without any need
of form factors) the extended, nonspherical and deformable character of
nucleons and the alteration of their intrinsic magnetic moment while
preserving
the conventional orbital and intrinsic angular momenta 	and other physiacal
laws}.
These conditions are necessary for consistency, evidently because neutrons are
under {\sl external} electromagnetic fields for which the angular momenta are
preserved.

It is important to apply Eq.s (4.8) and (4.9), first, to the exact
representation of $4\pi$-interferometric measures of type (1.3), and then to
the exact representation  of total magnetic
moments of few body nuclei. These results were first presented by Santilli
[15] during the meeting <Deuteron 1993>  at the JINR in Dubna, Russia.
However,
the calculations were done for values of $n_4\neq 1$ (interpreted as the
density of the neutrons), and  under a joint mutation of spin.
It is important to review these results for the more appropriate
interpretation
of ${n_4}$ introduced in this paper.

Assume that the $4\pi$-mutation is 1\%, yielding $\theta=713^{\circ}$,
which is of the order of magnitude of the measures (1.3).
The isotopies re-construct the exact $SU(2)$ symmetry in isospace,
thus requiring
$\h{\theta}=\theta /n_1\tim n_2=720^{\circ}$ [6g,6k]. This yields

$$
{n_1}^2={n_2}^2=713^{\circ}/720^{\circ}=
0.990\:,\quad{n_3}^2=1/0.990\tim 0.990
=1.020\:,\eqno{(4.10)}
$$

\no namely, the deformation is given by the transition from a perfectly
spherical charge disribution to one of {\sl prolate} character, exactly as
needed for the deuteron (Sect. 1).

Note that the different normalization ${n_1}^2+{n_2}^2+{n_3}^2=3$ (Sect. 2)
yields the values

$$
{n_1}^2=-{n_2}^2=0.990\:,\quad{n_3}^2=3-2\tim 0.990=1.020\:,\eqno{(4.11)}
$$
\no which coincide with values (4.10).

Then, assuming in first approximation that $\h{\mu}/\mu=n_4/n_3\approx
713^{\circ}/720^{\circ}$, we have the remaining value

$$
n_4=n_3\tim 713^{\circ}/714^{\circ}=1.000\:,\eqno{(4.12)}
$$

\no namely, the isodirac equation is capable of {\sl deriving} the value
$n_4=1$ occurring for motion in vacuum, exactly as it is the case for the
thermal neutron beam of tests [3].

We now study the application of isodirac equation (4.3) for the
exact-numerical
representation of the total magnetic moments of few-body nuclei. For this we
assume to a good approximation that protons and neutron have the same size and
shape, thus admitting the same ellipsoidical shape with
${n_1}^2={n_2}^2\l {n_3}^2$
or $\r {n_3}^2$, ${n_1}^2\tim{n_2}^2\tim{n_3}^2=1$.

In regard to the value of $n_4$, the motion of the proton and neutron in the
deuteron,
strictly speaking, {\sl is not} in vacuum because each particle is moving
within the wavepackets of the other, thus resulting in a difference of $n_4$
from 1. However the deuteron size is considerably bigger than the
nucleon charge
radius. As a result, we can assume {\sl in first approximation} that $n_4=1$,
with the understanding that a refinement of the data is expected whenever
{\sl experimental} information on the value of $n_4$ for the
deuteron is known.

A simple isotopy of the conventional QM model (see, e.g, [1]) then yields the
following {\sl isotopic theory for the total nuclear magnetic moments}

$$
\h{\mu}^{\tot}=\sum_k({{\h g}_k}^{\:\:L}\tim\h M_{k3}+
{{\h g}_k}^{\:\:s}\tim\h S_{k3})\:,
\eqno{(4.13a)}
$$

$$
\h g_n=g_nn_4/n_3\approx g_n/n_3\:,\quad\h g_p=g_p n_4/n_3\approx g_p/n_3\:,
\eqno{(4.13b)}
$$

$$
e\hbar /2m_pc_0=1\:,\quad {g_n}^s=-3.816\:,\quad{g_p}^s=5.585\:,
\eqno{(4.13c)}
$$

$$
{g_n}^M=0\:,\quad{g_p}^M=1\:,\eqno{(4.13d)}
$$

For the case of the deuteron, the above model yields the {\sl exact
representation of} ${\mu_D}^{\mbox{\scriptsize\rm exp}}$, Eq. (1.1),

$$
{\h{\mu}}_{\theor}^{\quad\tot}=g_pn_{4p}/n_{3p}+g_n n_{4n}/n_{3n}\approx
(g_p+g_n)n_4/n_3\equiv
{\mu_D}^{\mbox{\scriptsize\rm exp}}=0.857\:,\eqno{(4.14a)}
$$

$$
n_4=1.000\:,\quad n_3=1.000\tim 0.880/0.857=1.026\:,\eqno{(4.14b)}
$$

\no with consequential ellipsoidical shape of the two nucleons

$$
{n_3}^2=1.054\:,\quad{n_1}^2={n_2}^2=(1/{n_3}^2)^{1/2}=0.974\:.\eqno{(4.15)}
$$

\no which is precisely of the {\sl prolate} character, as expected (Sect. 1).

\vspace{1cm}

FIGURE 7. {\bf The structure of the deuteron according to relativistic
hadronic mechanics.} A schematic view of the  structure of the deuteron
according to the iso-Dirac equation (4.3) for which the charge distribution
of the individual nucleons is deformed into a spheroidal
ellipsoid of a {\sl prolate} type, which implies a
{\sl decrease} of the conventional values of the magnetic
moments of the individual nucleons when in vacuum. In turn, such a decrease
permits the apparently first exact-numerical representation of the total
magnetic moment of the deuteron. It should be indicated that the deformation,
expressed by Eqs. (4.15), is only of a few percentage points. Yet its
conceptual, theoretical and experimental implications are far reaching, as
indicated in the final part of this paper.

\vspace{1cm}

Note that the different normalization ${n_1}^2+{n_2}^2+{n_3}^2=3$ would
yield the values

$$
{n_3}^2=1.054\:,\quad{n_1}^2={n_2}^2=(3-{n_3}^2)/2=0.973\eqno{(4.16)}
$$

\no which are very close to the preceding ones.

Note that the data for the $4\pi$ spinorial symmetry tests, Eq.s (4.10), and
those for the deuteron, Eq.s (4.15), are very close. This illustrates that
the $4\pi$-interferometric measures, even though not inclusive of strong
nuclear forces, could provided experimental evidence on the alterability of
the intrinsic magnetic moments of nucleons in the deuteron structure and,
therefore, resolve the problem of total nuclear magnetic moments.

Note also that, along the historical hypothesis of Fig. 1, the fit (4.14) is
reached via a geometrical representation of the deformation of the charge
distribution, which is applicable to any preferred structure model, the
opposite approach of adapting the deformation of magnetic moments to any
conjectural structural model being manifestly questionable.

We finally note that the representation (4.14) is exact via an isorelativistic
treatment in first quantization based on only the $D$ state, while the
conventional relativistic treatment [2] uses all possible $S$-, $D$- and
$P$-states without achieving such an exact representation.

The following additional applications of HM and its relativistic extension
should also be indicated (in addition to those of ref.s [7]):

{\bf 1) Nuclear physics:} Reconstruction of the exact rotational symmetry for
deformed-oscillating nuclei [6k]; reconstruction of the exact isospin symmetry
in nuclear physics [6h]; axiomatically consistent representation of
dissipative nuclear processes [6k]; and others.

{\bf 2) Particle Physics:} reconstruction of the exact Minkowski space,
Poincar\'e symmetry and special relativity at the isotopic level [6k] for all
possible signature-preserving alteration of the flat space-time geometry [16];
exact-numerical representation of the behaviour of the meanlives of unstable
hadron with energy [17]; exact iso-Minkowskian representation [17a] and
experimental fit [18b] of the Bose-Einstein correlation [18c] for high energy
[18d] and low energy [18e] from first axiomatic principles and without
{\sl ad hoc and unknown} parameters as originating form the nonlocal-integral
interactions due to mutual overlapping of the wavepackets; numerical
interpretation of the synthesis of the neutron from protons and electrons only
as occurring in stars [6i]; isoquark theory [19] with conventional quantum
numbers, exact confinement and convergent perturbative series; reconstruction
of the  exact parity under weak interactions [6k]; new classical and quantum
theory of antiparticles characterized by the antiautomorphic isodual map of
conventional classical and quantum theories of matter, $A\arr A^d=-A^{\dag}$
[20]; and others.

{\bf 3) Gravitation and astrophysics:} Achievement of the universal invariance
of gravitation [6g] which is given by the isopoincar\'e symmetry of Sect. 3
for the particular case when the  iso-Minkowskian metric equals the Riemannian
metric, $\h{\eta}(x,\dot x,\ldots)=g(x)$; achievement of an operator form of
gravity which is as axiomatically consistent as RQM [21], which is again given
by the  isopoincar\'e formulations of Sect. 3 with $\h{\eta}=g(x)$, since they
are of operator character; numerical representation of the large difference in
cosmological redshift of quasars when physically attached to their associated
galaxies [22a,22b] as due to the decrease of the speed of light within the
huge quasars chromosphere represented via the iso-Minkowskian geometry under
the isospecial relativity; numerical representation of the internal quasars
redshift and blueshift [22c]; new isoselfdual cosmology with equal
distribution of matter and antimatter and total null characteristics of the
universe [6k]; and others;

{\bf 4) Superconductivity:} achievement of an explicitly {\sl attractive}
interaction among the two {\sl identical} electrons of the Cooper pair [23],
which is reached via the isotopic lifting of the conventional Coulomb problem
outlined in Sect. 3, in excellent agreement with experimental data.

{\bf 5) Theoretical Biology:} Axiomatic representation of irreversible and
nonconservative characters of biological systems; identification of the
apparent origin of irreversibility at the ultimate level of constituents with
nonlocal correlation effects; new geometric representations of locomotion and
bifurcations in biological systems; and others [24].

\section{Apparent new recycling of nuclear waste}

The experimental finalization of the alterability of the intrinsic magnetic
moments of nucleons via total nuclear magnetic moments, $4\pi$-interferometric
measures, or other means signals a necessary departure from the conventional
linear, local-differential and Hamiltonian-unitary formulation of the
Poincar\'e symmetry.

First, the above occurrence would establish the applicability in the nuclear
structure of the isopoincar\'e symmetry, the isospecial relativity and related
RHM. Since the latter are {\sl directly universal} for all  possible
alterations of the geometry of empty space, {\sl they would then apply even
when not desired}.

At any rate, the isopoincar\'e symmetry is the only generalized symmetry known
to this author which permits the preservation of the abstract axioms of the
special relativity under nonunitary maps, because conventional deformations
[4] imply necessary structural departures. As a matter of fact, the abstract
identity  of the isotopic and conventional symmetries,
$\h P(3.1)\equiv P(3.1)$, with consequential preservation under ``isotopic
completion" of conventional physical laws explains the reasons why their
experimental validity, by no means, implies that conventional QM is the only
applicable theory.

Once the above elements are understood, implications of the historical
hypothesis of Fig. 1 are consequential. The fist consists of apparently new
means for recycling nuclear waste which can be used by the nuclear power
companies in their own plants, thus avoiding altogether the dangerous and
expensive transportation of the waste to yet un-identified dumping site.

In essence, RHM in its nuclear realization and the fundamental isopoincar\'e
symmetry predict the possible mutation not only of the intrinsic magnetic
moment of the neutron, but also of its meanlife, to such an extent that the
former implies the latter and viceversa (as one can see via the use of the
isoboosts). In turn, the  control of the meanlife of the neutron {\sl de
facto} implies new means for recycling the nuclear waste.

In this respect, the first physical reality which should be noted (and
admitted) is that total nuclear magnetic moments constitute {\sl experimental
evidence} on the  alterability of the inrinsic magnetic moments of nucleons.

The second physical reality which should be noted (and admitted) is that, by
no means, the neutron has a constant and universal meanlife, because it
possesses
a meanlife depending on the local conditions. In fact, the neutron's meanlife
is of the order of seconds when belonging to certain nuclei with rapid beta
decays; a meanlife of the order of 15 minutes when in vacuum; a meanlife of
the order of days, weeks and years when belonging to other nuclei; all the way
to an infinite meanlife for stable nuclei.

Once the above occurrences are admitted, the {\sl basic principle for possible
new recycling of nuclear waste is the ``stimulated neutron decay" (SND)
consisting of resonating or other subnuclear mechanisms suitable to stimulate
its beta decay}. Among the various possibilities under study, we quote here
the possible {\sl gamma stimulated neutron decay}. (GSND) according to the
reaction [25a]

$$
\gamma +n\arr p^++e^-+\bar{\nu}\:,\eqno{(5.1)}
$$

\no which is predicted by RQM to have a very small (and therefore practically
insignificant) cross section as a function of the energy, but which is instead
predicted by RHM to have a resonating peak in the cross section at the value
of 1.294 MeV (corresponding to $3.129\tim 10^{20}$ Hz). As such, the above
mechanism is of {\sl subnuclear} character, in the sense of occurring in the
{\sl structure of the neutron}, rather than in the nuclear structure, the
latter merely implying possible refinements of the resonating frequency due to
the (relatively smaller) nuclear binding energy [7b].

When stable elements are considered, the above GSND is admitted only in
certain instances, evidently when the transition is compatible with all
conventional laws. This is the  case for the isotope Mo(100,42) which, under
the GSND, would transform via beta emission into the Te(100,43) which, in
turn, is naturally unstable and beta decays into Ru(100,44). For a number of
additional admissible elements see [25a].

The point important for this note is that the GSND is predicted to be
admissible for large and unstable nuclei as occurring in the nuclear waste.
The possible new form of recycling submitted for study in this note is given
by  {\sl bombarding the radioactive waste with a beam of photons of the needed
excitation frequency and of the maximal possible intensity}. Such a beam
would cause an instantaneous excess of peripheral protons in the waste nuclei
with their consequential decay due to  instantaneous excess of Coulomb
repulsive forces.

It should be stressed that this note can only address the basic
{\sl principle} of the GSND. Once experimentally established (see later on),
the recycling requires evident additional technological studies on the
{\sl equipment} suitable to produce the photon beam in the desired frequency
and intensity, e.g., via synchrotron radiation or other mechanisms [25c].

The imporatant point is that equipment of the above nature is expected to be
definitely smaller in size, weight and cost than large particle accelerators.
As such, the recycling is expected to verify the basic requirement of
usability by the nuclear power companies in their own plants.

A novelty of the proposed new recycling is that  it  is specifically conceived
to   occur at the {\sl subnuclear} level. A virtually endless number of
possibilities exist for the  reduction of the meanlife of the  waste via
mechanisms of {\sl nuclear} type. Among them we note mechanisms based on RQM,
such as those by Shaffer et al. [26a], Marriot et al. [26b], Barker [26c] and
others, as well as new {\sl nuclear} mechanisms predicted by RHM and currently
under patenting. The understanding is that, to maximize the efficiency, the
final equipment is expected to be  a combination various means of both
subnuclear and nuclear character.


\section{Possible additional advances}

The possible experimental verification of the alterability of the magnetic
moment and meanlife of the neutron would rather deep implications throughout
all aspects of nuclear physics. In addition to possible new forms of
recycling nuclear waste indicated above, it may be of some value to indicate
the following additional, possibilities.

\vspace{0.5\baselineskip}

{\bf 1) Nuclear forces.} RHM terminates the study of nuclear forces by
adding terms and terms in the Hamiltonian, because it provides means for
rigorous new studies based on the representation with potentials of terms
truly being of  action-at-a-distance, and the representation with the isounit
of contact, nonlinear, nonlocal and nonunitary effects for which the notion
of a potential has no conceptual, mathematical or physical meaning of any
nature;

\vspace{0.5\baselineskip}

{\bf 2) Nuclear structure.} RHM permits a deeper understanding of the nuclear
structure via the admission of small, yet significant interactions of {\sl
nonlocal-integral and nonhamiltonian-nonunitary} type, due to the
wave-overlapping of the constituents. In turn, it can be safely stated that
the inclusion of the latter interactions will inevitably lead to new nuclear
models. New  nuclear reactions cannot also be excluded in view of the
{\sl attractive} character of the latter interactions under certain rather
selective but well identified conditions even against the Coulomb barrier, as
theoretically and experimentally established in [23].

\vspace{0.5\baselineskip}

{\bf 3) Controlled fusion.} One of the first applications of the studies of
this paper is expected for the attempts at reaching a ``hot controlled fusion"
with a positive energy balance. In fact, these attempts are essentially
costituted by a condensation phase caused by magnetic or other means which
works as predicted, followed by the initiation of the fusion process with its
notorious instabilities which have not been controlled to date.

It is evident that RQM is exactly valid for the condensation phase due to
large mutual distances. However, the exact validity of the same discipline at
the initiation of the fusion process does not appear to have solid
scientific grounds because of the  emergence of numerous new effects at short
distances, such as those of nonlinear, nonlocal and nonpotential type, which
are beyond any hope of quantitative treatment via RQM. In view of evident
societal aspects, deeper studies with a covering disciplines appear,
therefore, to be warranted.

In particular, the studies of this paper suggest that {\sl the intrinsic
magnetic moments of protons and neutrons is expected to change precisely at
the initiation of the fusion process}. It is then evident, although not widely
admitted, that such  a change has implications for the currently uncontrolled
instabilities, and should be reflected in a re-design of the magnetic and
other confinement. After all, the control of the instabilities is currently
conducted under the (tacit) assumption that protons and neutrons preserve
their intrinsic magnetic process during the fusion process.

\vspace{0.5\baselineskip}

{\bf 4) New subnuclear energy.} Any new recycling of nuclear waste is
unavoidably linked to possible new sources of energy. In fact, the GSND

$$
\gamma_{\res}+\mbox{\rm Mo}(100,42)\arr_{\stim}\mbox{\rm Tc}(100,43)
+\beta\arr_{\spont}\mbox{\rm Ru}(100,44)+\beta\:,\eqno{(6.1)}
$$

\no is {\sl de facto} a potential new source of {\sl subnuclear}
energy called {\it hadronic energy} [25a] (see also the review [25b]) which
releases the rather large amount of about 5 MeV plus the energy would not
release harmful radiations, would not imply  radioactive waste, would not
require heavy shield or critical mass, and would be realizable in large or
minuterized forms.

\vspace{0.5\baselineskip}

{\bf 5) Other possible applications.} Numerous additional applications are
conceivable, such as the production of rare isotopes via GSNT, the prediction
of neutron rays for industrial applications via the synthesis of the neutron
from protons and electrons beams in flight, medical applications, and others.


\section{Proposed experiments}

The continuation of quantitative scientific studies on the proposed new
recycling of the nuclear waste (as well as on the other applications
indicated above) beyond the level of personal views one way or another,
requires the following three basic experiments, all of truly fundamental
character, moderate cost and full realization with current technology.

\vspace{0.5\baselineskip}

{\bf I. Finalize the interferometric $4\pi$ spinorial symmetry measures [3].}
The above measures are manifestly fundamental for possible new forms of
recycling
as well as for possible new forms of subnuclear energy. In fact, they would
provide experimental evidence on possible deviations from the Poincar\'e
symmetry in favor of our covering isopoincar\'e form. This is due to the fact
that, if confirmed, the measures would establish a deviation from the
fundamental {\sl spinorial} transformation law in favor of the mutated form
(4.8). The alterability of the meanlife of the neutron would then be
consequential.

It should be stressed that the primary evidence for the alterability of
the intrinsic magnetic moments of nucleons rests in the experimental values
of total nuclear magnetic moments. The finalization of interferometric
measures [3] would merely provide additional backing on the same alterability.
The latter would however occur for controllable conditions, thus being
invaluable for other predictions via extrapolations.

\vspace{0.5\baselineskip}

{\bf II. Repeat don Borghi's experiment [25d] on the apparent synthesis of the
neutron from protons and electrons {\bf\sl only}.} Despite momentous
advances, we
still miss fundamental experimental knowledge on the  structure of the
neutron,  e.g., on how the neutron is synthesized from protons and electrons
{\sl only} in young stars solely composed of hydrogen (where quark models
cannot be used owing  to the lack of the remaining members of the baryonic
octet, and weak interactions do not provide sufficient information on the
{\sl structure} problem).

The synthesis occurs according to the reaction

$$
p^++e^-\arr n+\nu\:,\eqno{(7.1)}
$$

\no  which: is the ``inverse" of the stimulated decay (6.1); is predicted by
RQM to have a very small cross section as a function of the energy; while the
same cross section is predicted by RHM to have a peak at the threshold energy
of 0.80 MeV in singlet $p$-e coupling [6i].

The possible synthesis of the neutron has a fundamental relevance for waste
recycling, besides other industrial applications. If the electron
``disappears" at the creation of the neutron, as in current theoretical views,
the GSND becomes of difficult if not impossible realization.

However, the electron is a permanent and stable particle. As such, doubts as
to whether it can ``disappears" date back to Rutherford's very conception of
the neutron as a ``compressed hydrogen atom". As well known,
RQM does not permit
such a representation of the neutron structure on numerous counts.
Nevertheless, the covering RHM has indeed achieved an exact-numerical
representation of {\sl all} characteristics of the neutron according to
Rutherford's original conception [6i].

The novelty permitting the above result is that, when immersed within the
hyperdense proton, the electron experiences a mutation $e^-\arr\h e^-$ of its
{\sl intrinsic} characteristics (becoming a quark ?) including its rest energy
(because $n_4\neq 1$ inside the proton, thus $E_{\h e}=mc^2=
m_e{c_0}^2/{n_4}^2$). The excitation energy of 1.294 MeV is predicted by our
covering isopoincar\'e symmetry under the condition of recovering all
characteristics of the neutron for the  model $n=({p^+}_{\uparrow},
{{\h e}^-}_{\:\:\downarrow})_{\mb{\scriptsize\rm{RHM}}}$, including its
primary decay for which [6i] $\h e^-\arr e^-+\bar{\nu}$.

A preliminary experimental verification of the synthesize the neutron in
laboratory was done by don Borghi's and his associates [25d]. Since
experiments can be confirmed or dismissed {\sl solely} via other experiments
and certainly not via theoretical beliefs one way or the other, don Borghi's
experiment must be run again and either proved or disproved. The test can be
repeated either as originally done [loc. cit.], or in a number of alternative
ways, e.g., by hitting with a cathodic ray of 0.80 MeV a mass of beryllium
saturated with hydrogen, put at low temperature and subjected to an intense
electric field to maximize the $p$-e singlet coupling. The detection of
neutrons emanating from such a set-up would establish their synthesis.

\vspace{0.5\baselineskip}

{\bf III. Complete Tsagas' experiment [25e] on the stimulated neutron decay.}
A most fundamental information needed for additional quantitative studies is
the verification or disproof of the GSND at the resonating gamma frequency of
1.294 MeV.

The latter experiment has been recently initiated by N. Tsagas and his
associates [25e]. It consists of disk of the radioisotope $\mbox{Eu}^{152}$
(which naturally emits gammas of 1.3 MeV) placed parallel and close to a disk
of an element admitting of the GSND, such as the Mo(100, 42) (or a sample of
nuclear waste). The detection of electrons with at least 2 MeV emanating from
the system would establish the {\sl principle} of the GSND (because such
electrons can only be of subnuclear origin, Compton electrons being of at
most 1 MeV). The detection via mass spectrography of traces of the extremely
rare Ru(100,44) after sufficient running time would confirm said principle
beyond any reasonable doubt. The practical realization of the proposed form
of waste recycling would then be shifted to the  industrial development and
production of a photon beam of the needed frequency and intensity via a
relatively small equipment usable by the nuclear power companies in their own
plants.

\vspace{1cm}

\end{document}